\documentclass{aa}
\usepackage{times}
\usepackage{graphicx}
\begin{document}

  \thesaurus{ 03.20.8 
              11.09.01 NGC~4214 
              11.09.4 
              11.09.5 
	      11.11.1 
              11.19.3 
	      }

\title{Kinematical analysis of the ionized gas in the nuclear region of 
NGC~4214}
\titlerunning{Kinematical analysis of the ionized gas in NGC~4214}

\subtitle{}

\author{J. Ma\'{\i}z-Apell\'aniz\inst{1,2} 
   \and C. Mu\~noz-Tu\~n\'on\inst{3}
   \and G. Tenorio-Tagle\inst{4} 
   \and J. M. Mas-Hesse\inst{1}}

\offprints{J. Ma\'{\i}z-Apell\'aniz - jma@laeff.esa.es}

        \institute{Laboratorio de Astrof\'{\i}sica Espacial y F\'{\i}sica
                   Fundamental - INTA, POB 50727, E-28080 Madrid, Spain.
           \and Departamento de Matem\'aticas y F\'{\i}sica Aplicada,
                   Universidad Alfonso X el Sabio,
                   E-28691 Villanueva de la Ca\~nada, Madrid, Spain.
           \and Instituto de Astrof\'{\i}sica de Canarias, E-38200 La Laguna,
                   Tenerife, Spain.
           \and Instituto Nacional de Astrof\'{\i}sica, \'Optica y 
		   Electr\'onica, Apartado Postal 51, 72000 Puebla, M\'exico.
           }

\date{Accepted 2 December 1998}

\maketitle

\begin{abstract}

We present in this paper a detailed study of the kinematical properties of
the ionized gas around the young massive star clusters in the nucleus of
NGC~4214. The analysis is based on bidimensional spectroscopical data,
allowing to derive the spatial variation of different properties
(intensity, velocity and width~/~line splitting) of the emission lines 
H$\alpha$ and [O\,{\sc iii}]~$\lambda$5007 along the nuclear region. We have 
found that the Giant H\,{\sc ii}
region around the two most massive clusters in NGC~4214 (A and B) is
resolved into two clearly separated regions. We have not detected
superbubbles with the properties we would expect according to the
evolutionary state of the stellar clusters, but just a partial ring feature
around the most massive one and two expanding shells around cluster B. The
first expanding shell seems to have experienced blowout, whereas the second one 
is still 
complete. A possible explanation to this phenomenon is that the most massive 
stars in a starburst spend a large fraction of their lives buried inside their 
original molecular clouds.  Champagne flows might have formed at the borders of 
the regions, especially on the SE complex, explaining the existence of the
diffuse ionized gas around the galaxy. As a consequence of these results we
postulate that NGC~4214 is indeed a dwarf spiral galaxy, with a thin ($\sim
200$ pc) disk that inhibits the formation of large scale structures in the
ISM. The mechanical input deposited by the star formation complexes, in a
variety of physical processes that include the free expanding bubbles
liberated after blowout and photoevaporation of the parent clouds, have
succeeded in generating the structures now detected far from the disk,
giving place to the large-scale structure which now enriches the optical
appearance of the galaxy.

\end{abstract}

\section{Introduction}

\hyphenation{GHIIRs}

Magellanic irregular galaxies are excellent laboratories for the study of
massive star formation and its impact on the surrounding interstellar
medium due to the large number of Giant H\,{\sc ii} Regions (GHIIRs) that they
host. For example, more than 40 GEHRs have been studied and catalogued
along the central bar of NGC~4449 (Mu\~noz-Tu\~n\'on
et al. 1998).  Magellanic irregulars are also particularly rich in gas (Hunter
\& Gallagher 1990) and usually present an extended emission of
Diffuse Ionized Gas (DIG), most probably ionized by photons escaping from the
GHIIRs (Mu\~noz-Tu\~n\'on et al. 1998).  Also, superimposed on the DIG,
kiloparsec-scale filaments, bubbles, and shells can be found (termed as 
``froth'' by Hunter\& Gallagher (1990)) and are most probably remnants
of past star formation episodes. Further evidence supporting
the idea of one or more past star formation episodes rests on the presence of
Faint Extended Broad Emission Lines (FEBELs) detected on some of these
irregular galaxies.  The presence of FEBELs is probably linked to the low
metallicity of these objects (Tenorio-Tagle et al. 1997).

In this paper we deal with NGC~4214. This galaxy has been
classified as SBmIII by Sandage \& Bedke (1985), but is frequently
considered a typical Magellanic irregular. It shows important star formation
activity along a bar-like structure across its nucleus. In a
previous article (Ma\'{\i}z-Apell\'aniz et al. 1998, hereafter Paper I) we
used bidimensional spectroscopy to study the excitation and density of the
gas as well as its distribution relative to the stars and dust.  In Paper I
we described the existence of two star-forming regions: the larger
($\sim$13,000 OB stars), slightly older (3.0-3.5 Myr) NW complex and the
smaller ($\sim$4,000 OB stars), slightly younger SE complex.  The NW
complex is often named in the literature NGC 4214-I, and the SE complex,
NGC 4214-II.
The age of each complex was determined comparing four different parameters
($W($H$\beta)$, $W$(WR), $I({\rm WR bump})/I($H$\beta$), and $T_{eff}$) with 
the predictions of synthesis models
(Cervi\~no \& Mas-Hesse 1994).  The consistency of our age determination
using different methods indicates that the number of ionizing photons
destroyed by dust or escaping from the star forming region might be large
but less than half of the total ($\approx$ 30\%).  The slightly younger age
and smaller size of the SE complex is consistent with its morphology: three
continuum knots with coincident H$\alpha$ emission maxima surrounded by
dusty clouds.  On the other hand, the NW complex displays a more
complicated structure.  It shows two main continuum knots surrounded by
several H$\alpha$ knots and an extended halo with dust detected only at its
borders.  Also visible at the NW complex are two H$\alpha$ minima located close
to each of the two continuum knots.

In this paper we continue our study of NGC~4214 analyzing the kinematical
properties of the ionized gas using bidimensional spectroscopy.  The
spatial variations in intensity, velocity and velocity dispersion
($\sigma$) and the presence of anomalous kinematical components in the gas
emission, in combination with the results of Paper I, allow us to produce
an overall picture of the starbursts in NGC~4214. In Section 2 we present
our observations and explain the reduction procedure. The results are
presented in Section 3 and discussed in Section 4.

\section{Observations and Data Analysis}

Optical long-slit spectra of NGC 4214 were obtained on April 19-20/1992
with the ISIS spectrograph of the William Herschel Telescope, as described
in Paper~I.  The spectra were obtained at twelve parallel positions, all of
them at a P.A.  of 150\hbox{$^\circ$}, which almost coincides with the galaxy 
minor
axis as well as with the orientation of the present day star-forming
regions along the bar-like structure.  The effective width for each slit
was 1\hbox{$^{\prime\prime}$} and the spacing between the centers of 
consecutive positions
was set equal to 2\hbox{$^{\prime\prime}$} (see Fig.~1 in Paper I).

Two spectra were taken simultaneously at each position, one in the range
between 6390~and~6840~\AA\ (red arm) and the other one in the range between
4665~and~5065~\AA\ (blue arm), both of them with a dispersion of approximately
0.4~\AA /pixel, with the intention of studying H$\alpha$ and 
[O\,{\sc iii}]~$\lambda$5007 simultaneously.
The spatial resolution along the slit was 0\hbox{$.\!\!^{\prime\prime}$}34/pixel
and 0\hbox{$.\!\!^{\prime\prime}$}36/pixel
for the red and blue spectra, respectively.  The seeing varied between
0\hbox{$.\!\!^{\prime\prime}$}7 and 1\hbox{$.\!\!^{\prime\prime}$}0. This 
observational setup produces a three dimensional
(2D spatial + 1D frequency) data set that can be arranged as a series of high
dispersion spectra associated to a regular array of points in a map.  The $x$
axis runs along the slit from NW to SE and the $y$ axis increases with slit
number, from NE to SW.  The origin is fixed in such a way that the 
first point in each slit is centered at $x=1/3\hbox{$^{\prime\prime}$}$ and the 
first slit is 
centered at $y=2\hbox{$^{\prime\prime}$}$ (thus, the last slit is centered at
$y=24\hbox{$^{\prime\prime}$}$).

The data reduction was done using standard IRAF packages and our own procedures
at the LAEFF SUN-Unix environment.  Both atmospheric emission lines and 
reference neon and copper lamps were used to calibrate our spectra in
wavelength. The combined data lead to an estimation of the absolute errors in
the wavelength calibration of 0.04~\AA\ and 0.08~\AA\ for the red and blue
spectra, respectively.

The instrumental width of the lines, \mbox{$\sigma_{\rm inst}$}, was 
$11.6\pm 0.8$ km s$^{-1}$ and
$20.9\pm 0.8$ km s$^{-1}$ for the red and blue spectra, respectively.  The
lower instrumental width combined with the longer wavelength for a similar
dispersion leads to a better resolution in velocity for the red spectra when
compared to the blue spectra.  As a consequence of this, many points in the red
spectra which in H$\alpha$ showed non-single-Gaussian profiles (i.e. which 
showed multiple velocity components) appeared blended into a single Gaussian, 
or almost Gaussian, profile in [O\,{\sc iii}]~$\lambda$5007. 
Visible multiple components were quite scarce in
[O\,{\sc iii}]~$\lambda$5007 and this dearth allowed us to extract the 
[O\,{\sc iii}]~$\lambda$5007 kinematical
information by fitting a single Gaussian profile for each point.  This was
done using the automatic {\em fitlines} package developed by Jos\'e Acosta
at the Instituto de Astrof\'{\i}sica de Canarias.  On the other hand, the
existence of multiple components in H$\alpha$ required a manual fitting for each
point with one to three Gaussians using the {\em splot} package.
Both for [O\,{\sc iii}]~$\lambda$5007 and H$\alpha$ a 3 pixel smoothing
in the spatial direction was performed before the (single or multiple)
Gaussian fitting in order to obtain a better signal-to-noise ratio.

This combined strategy produced values for the flux, $\sigma_{\rm obs}$
(observed width) and center for each of the measured Gaussian components (one
for [O\,{\sc iii}]~$\lambda$5007, one to three for H$\alpha$) at each point. 
From those values,
bidimensional tables of intensity (flux per unit solid angle), $\sigma$ (width
due to random motions, see Mu\~noz-Tu\~n\'on 1994) and central velocity $v$
were obtained.  The relation between the observed ($\sigma_{\rm obs}$) and the
adopted line width \mbox{$\sigma$} is given by:

\begin{equation}
\sigma^2 = \sigma_{\rm obs}^2 - \sigma_{\rm inst}^2 - \sigma_{\rm th}^2 ,
\end{equation}

\noindent where \mbox{$\sigma_{\rm th}$} is the thermal width. Following 
Kobulnicky \& Skillman
(1996), the temperature was taken to be $10\mbox{,}500\pm 500$ K 
in all observed regions.

Estimated errors were computed for $\sigma$ and $v$.  The error sources
considered for $\sigma$ were: Direct measurement error (provided by the 
{\em fitlines} package), uncertainty in the instrumental width and uncertainty 
in the temperature.  The error sources considered for $v$ were: Direct 
measurement error (provided by the {\em fitlines} package) and calibration 
error.  For most cases, the direct measurement error was not important.  
However, there are some positions where the strong blending of the lines 
increased it significantly.


\begin{figure*} 
  \centerline{\includegraphics*[angle=90,width=\linewidth]{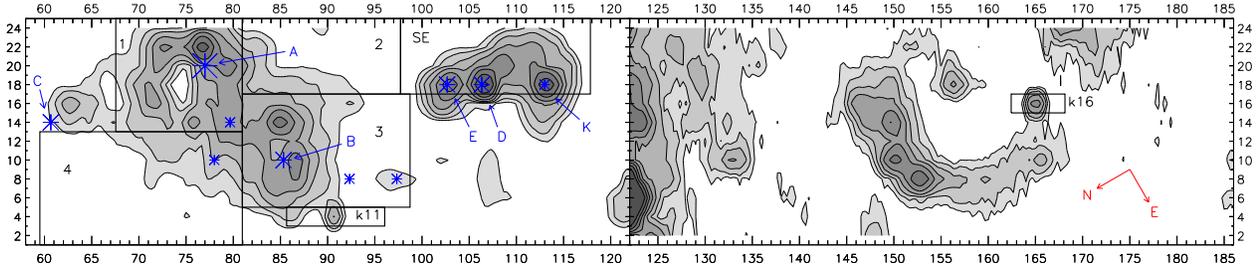}}
  \caption[]{H$\alpha$ synthetic map of NGC~4214 obtained from long-slit 
  spectra.
  The left side corresponds to the sites of most intense star formation and has
  a minimum contour value of 
  300$\cdot 10^{-17}$ erg s\mbox{$^{-1}$} cm\mbox{$^{-2}$} arcsec\mbox{$^{-2}$} 
  and a maximum of 7500 in these units, with the other six levels 
  logarithmically spaced. The right side, where H$\alpha$ emission is much 
  weaker, has 
  minimum and maximum values of 20 and 300 (note that the minimum contour on
  the left is the maximum contour on the right), with the other six levels 
  also logarithmically spaced. The axes are labelled in arcseconds with respect
  to the first pixel of the first slit (1\hbox{$^{\prime\prime}$} = 20 pc), 
  with $x$ in the
  horizontal direction and $y$ in the vertical direction (see text). This 
  coordinate system will be used throughout this work. The regions of interest 
  have been marked as 1, 2, 3, 4, k11 (altogether known as the NW complex or 
  NGC 4214-I), SE (the SE complex or NGC-4214-II) and k16. The asterisks 
  indicate the location of the continuum knots, the size being a measurement of 
  their intensity. The continuum knots referenced in the text are labelled as 
  A, B, C, D, E and K (see Paper I).}
\label{hacr0}
\end{figure*}

\section{The Gas Dynamics of NGC 4214}

Fig.~\ref{hacr0} shows the H$\alpha$ map of NGC 4214, which was already 
described in
Paper I.  Here we use it as a reference for the presentation and discussion of
our spectroscopic results and to indicate the main features of NGC 4214.  Big
asterisks indicate the location of the optical continuum maxima or the location
of the ionizing clusters (A,B,C following the nomenclature from Paper I).  The
two large H\,{\sc ii} regions lying along the optical bar -- which coincides 
with the
galaxy's minor axis as defined in the H\,{\sc i} map (McIntyre, 1998) -- are
clearly seen.  The largest one, NGC 4214-I (or NW complex), located around the 
optical-UV nucleus (Fanelli et al. 1997) is at horizontal coordinates 
$x$~=~60\hbox{$^{\prime\prime}$} to 
$x$~=~100\hbox{$^{\prime\prime}$} while the second one, NGC 4214-II (or SE 
complex), extends 
from $x$~=~100\hbox{$^{\prime\prime}$} to $x$~=~120\hbox{$^{\prime\prime}$} 
(see Fig.~\ref{hacr0}). Note that 
the right hand side of the map was drawn with a lower intensity threshold to 
unveil the much lower intensity emission features found at large distances from 
the galaxy nucleus.  Threshold values are indicated in the figure caption.

On the map we have marked several zones which have been analysed in detail.
The first one, zone 1, covers an area of 13\hbox{$^{\prime\prime}$} $\times$
12\hbox{$^{\prime\prime}$}
(equivalent to 260 pc $\times$ 240 pc) and includes the optical-UV nucleus of
the galaxy and the brightest H$\alpha$ emitting area of NGC 4214-I which
includes the bright H$\alpha$ knots 3, 7, 8, 9 and 10 (according to the 
nomenclature used in Paper I).

Zone 2 covers an area of 16\hbox{$^{\prime\prime}$} $\times$
8\hbox{$^{\prime\prime}$} (equivalent to 320 pc
$\times$ 160 pc) towards the southeastern border of zone 1 and spans the lower
intensity area at the optical SW edge of the bar, where Hunter \& Gallagher
(1990) detected the filamentary low emission structures or ``froth'' features,
emanating from the central regions of the galaxy.

Zone 3 covers a similar size area and contains most of NGC 4214-I not included
in zone 1; the bright H$\alpha$ knots 5 and 6 --see Paper I -- as well as 
continuum
knot B.  Zone 4 traces the background H$\alpha$ emission at the northern border 
of NGC 4214-I. Knot C was not included in any zone because it is an 
apparently older cluster which has almost no gas in its surroundings. The only 
remaining part of the original ionized cloud is located some 70 pc towards the 
south of the cluster. This cloud does not show any important kinematical
difference with its surroundings and, therefore, will not be studied in this
section.

We have also analyzed the spectra of the three main H$\alpha$ knots of NGC 
4214-II (zone SE) and that of two small
but intense and peculiar emitting knots.  One of these is on the border of NGC
4214-I (k11) and the second one (k16) is a low intensity emitting region
50\hbox{$^{\prime\prime}$} (1~kpc) away from NGC 4214-II.

The H$\alpha$ emission line for each of the spectra included in the marked 
areas has 
been analyzed by fitting either one, two or three Gaussian profiles and from 
these we derived intensity, velocity and $\sigma$ values (see the previous 
section). The parameters have been
plotted as a function of position along each of the slits that traverses the
marked zones.  All velocity dispersion ($\sigma$) values have been corrected
for instrumental and thermal broadening.  The velocity reference has been taken
from the H\,{\sc i} map (McIntyre, 1998).  From this map, the heliocentric 
velocity at the nucleus is 300 km s$^{-1}$.  This value has been substracted to 
all measured velocity centroids presented in this paper.

\subsection{NGC 4214-I (the NW complex)}

The largest star-forming complex, NGC 4214-I, is covered by zones 1, 2, 3 
and 4 (see Fig.~\ref{hacr0}) all of them presenting different gas dynamical 
properties.

\subsubsection{Zone 1}

\begin{figure*}
\centerline{\includegraphics*[width=\linewidth]{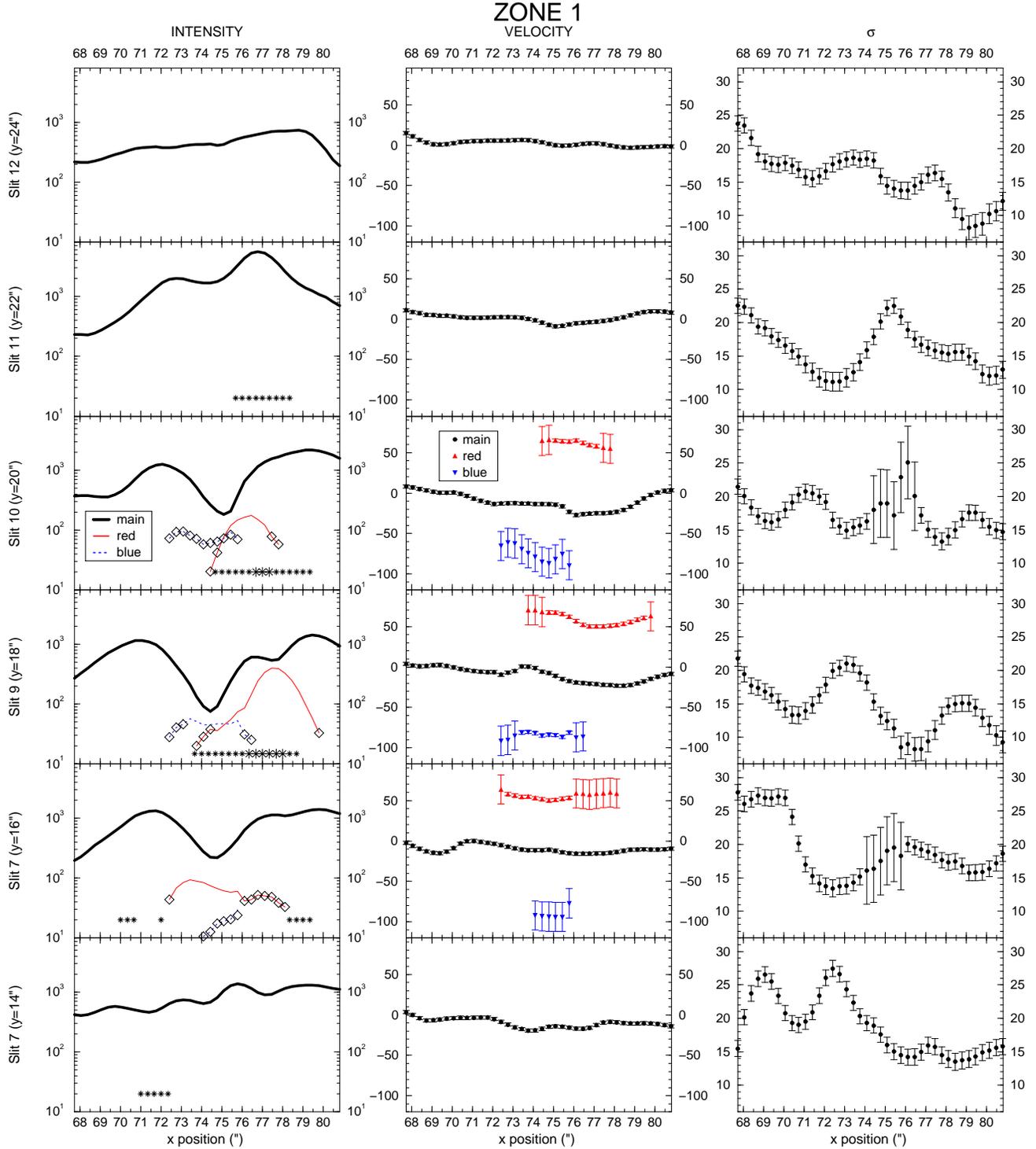}}
\caption{Intensity (first column), velocity (second column) and \mbox{$\sigma$} 
(third column) values fitted to the H$\alpha$ emission corresponding to the 
spectra 
of zone 1. In the intensity plots, the thicker line represents 
the main component. When the red or blue components are detected they are drawn
with a thin continuous line (red) or with a thin dashed line (blue). 
Large asterisks mark continuum knot A while the small ones trace the 
surrounding extended young population
(see Paper I). Diamonds indicate marginal detections as exemplified in 
Fig.~\ref{spec_zone1}. Different symbols are used in the velocity plots to
represent the main, blue and red components. The sigma is shown for the main
component only. The separation between consecutive slits is
2\hbox{$^{\prime\prime}$}
(40 pc), and the horizontal scale is in arcseconds (1\hbox{$^{\prime\prime}$} = 20 pc). The 
units are $10^{-17}$ erg s\mbox{$^{-1}$} cm\mbox{$^{-2}$} arcsec\mbox{$^{-2}$} for the intensity and km s\mbox{$^{-1}$}
for the velocity and \mbox{$\sigma$}. }
\label{zone1}
\end{figure*}

\begin{figure}
\centerline{\includegraphics*[width=\linewidth]{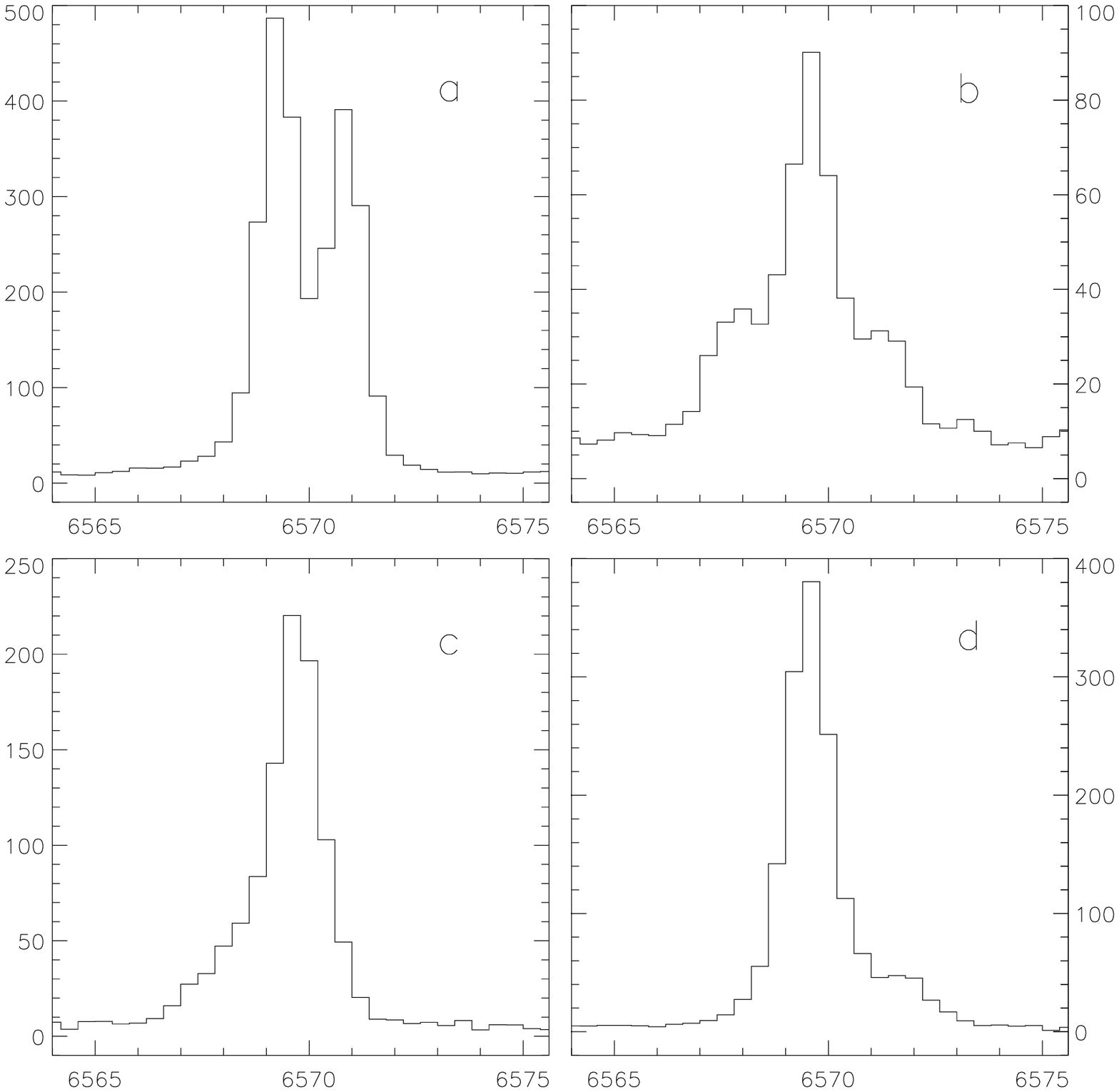}}
\caption{Examples of full and marginal detections. Four spectra taken at 
different locations along slit 9 are shown. The spectra show well resolved two 
and three components (a and b) as well as marginal detections (c and d). The 
units are \AA\ and  $10^{-17}$ erg s\mbox{$^{-1}$} 
\AA\mbox{$^{-1}$} cm\mbox{$^{-2}$} arcsec\mbox{$^{-2}$} for the $x$ 
and $y$ axes, respectively.}

\label{spec_zone1}
\end{figure}

Fig.~\ref{zone1} shows the results in H$\alpha$ corresponding to the
six slits (7 to 12) that traverse zone 1. At several 
spatial positions, up to three line emission components have been identified, 
and their intensities and velocities are also represented in the figures.  
Diamond symbols have been used to indicate marginal detections. 
As an example,  
Fig.~\ref{spec_zone1} shows four spectra extracted from 
slit 9. The spectrum in Fig.~\ref{spec_zone1}a shows a clear and well 
resolved two component splitting. Fig.~\ref{spec_zone1}b is a clear case with
three resolved components and Figs.~\ref{spec_zone1}c and d are 
examples of marginal detections: a blue secondary component 
in~\ref{spec_zone1}c and a red one in case~\ref{spec_zone1}d. No
attempt has been made to assign a $\sigma$ value to the secondary 
components. However, large $\sigma$ values of the main emission 
line may result from confusion with nearby unresolved red or 
blue secondary components, leading to possible fluctuations in $\sigma$ across 
the slit.

In zone 1 (see Fig.~\ref{zone1}) an extended, low emission, central valley is
surrounded by a broad rim which extends from 
$x$~=~71\hbox{$^{\prime\prime}$} to $x$~=~79\hbox{$^{\prime\prime}$} between 
slits 7 to 12. The 
full dimension of the ring-like structure is 200 pc $\times$ 200 pc 
and the H$\alpha$ intensity is a factor of 100 lower in the center or valley 
region. 

The low intensity H$\alpha$ valley is almost coincident with the location of the
young stellar cluster derived from the continuum map (Knot A) and is indicated
in the figure with large asterisks. The main emission line component is
present over the whole area and does not show large velocity variations over
zone 1, with a mean value about 0~$\pm$~10~km~s\mbox{$^{-1}$}, with respect to 
the systemic one.

Blue and red secondary components have been detected over an area of about
140 pc $\times$ 120 pc, extending over the ring-like structure albeit being
more evident on the H$\alpha$ depression and at the position of the young 
stellar
cluster.  The blue and the red secondary line components are displaced by about
70 km s$^{-1}$ from the galaxy bulk emission, presenting small velocity
variations along the slits ($\leq$10 km s\mbox{$^{-1}$}).

The velocity width of the main emission line component, ranges from about 10 km
s\mbox{$^{-1}$} to 28 km s\mbox{$^{-1}$}.  The weighted mean sigma value for t
he whole zone is 
supersonic and about 17 km s$^{-1}$. This value is only slightly larger
than the integrated one (15.0$\pm$0.6 km s\mbox{$^{-1}$}) over the whole NGC 
4214-I (zones
1-4) measured by Roy et al. (1986) and Arsenault \& Roy (1988).

On the other hand, the NW border of zone 1 traces the edge of the parent cloud,
where the line intensity rapidly drops by more than one order of magnitude (as
one goes from $x$=~72\hbox{$^{\prime\prime}$} towards
$x$=~68\hbox{$^{\prime\prime}$}, across slits 8 - 11),
and at the same time, the $\sigma$ value increases steadily to highly
supersonic values (20-30 km s$^{-1}$) while the velocity line center remains
well behaved.  An abrupt density drop is necessarily accompanied by a sharp line
intensity decrease and if this coincides with the region where the lines become
broader, then both observables seem to be tracing a champagne flow
(Tenorio-Tagle 1979) playing a major role at the boundary of the ionized cloud.

Although not indicated in Fig~\ref{zone1}, a low-level, very wide
($\sigma\approx 600-700$ km s\mbox{$^{-1}$}) H$\alpha$ component was also 
detected at the
location of the young stellar cluster.  This very wide component was
discovered by Sargent \& Fillipenko (1991), who attribute its origin to the
WR stars known to be present there.  Our data corroborates this broad
component is centered at the position of the cluster and is not present
anywhere else.

\subsubsection{Zone 2}

\begin{figure*}
\centerline{\includegraphics*[width=\linewidth]{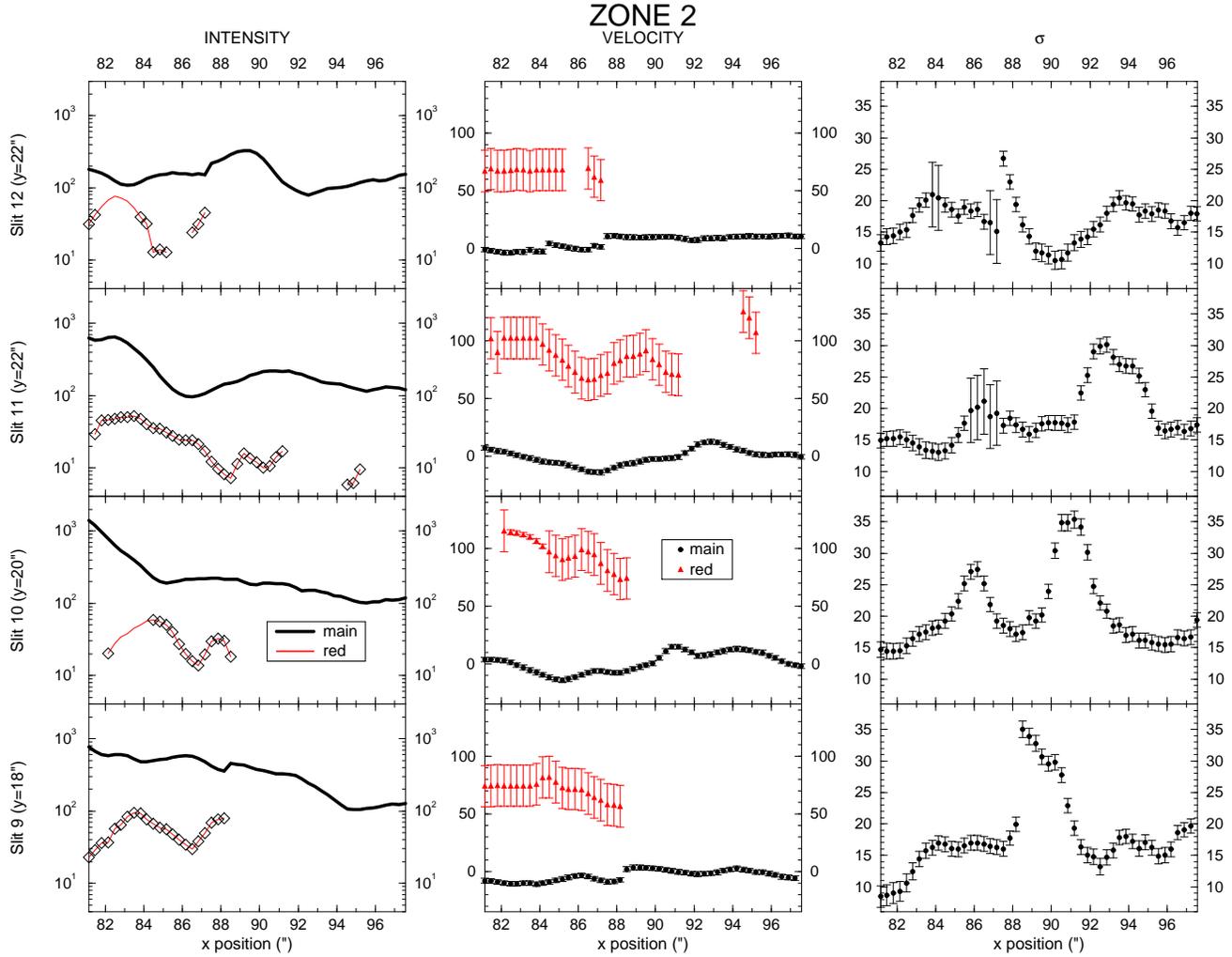}}
\caption{Intensity, velocity and \mbox{$\sigma$} values fitted to the H$\alpha$ 
emission 
corresponding to the spectra from zone 2. Symbols, lines, slit separation,
horizontal scale and units are as in Fig.~2.}

\label{zone2}
\end{figure*}

Fig.~\ref{zone2} shows the results from the detailed analysis of slits 9 - 12 
that traverse zone 2.   
Zone 2 extends from the SE border of zone 1. The main line intensity (thick 
line on Fig.~\ref{zone2}) decreases as
one moves towards the SE border of NGC 4214-I.  The line intensity falls by
almost one order of magnitude from position 81\hbox{$^{\prime\prime}$} to
95\hbox{$^{\prime\prime}$} on slits
9 to 11, over a distance of up to 280 pc, while the peak velocity of this main
emission component remains almost constant, at about 0 km s\mbox{$^{-1}$}, the 
same value measured on zone 1.

Looking at Fig.~\ref{zone2} in further detail one can see that the intensity
fall is more pronounced along the first 5 positions (100 pc) on slits 10 and
11.  At these coordinates a red secondary component is detected.  Since its 
S/N ratio is not sufficient for a secure fit to be carried out, no $\sigma$
values are given.  The red component velocity however is clearly shifted in a
range between 60-100 km s\mbox{$^{-1}$} from the bulk of the H$\alpha$ 
emission.  As in
Fig.~\ref{zone1} variations in $\sigma$ might reflect line contamination by
non-resolved components affecting the measured line width.  It seems that,
after reaching the edge of NGC 4214-I, which could be defined at 
$x =86$\hbox{$^{\prime\prime}$}, the ISM is disrupted, most probably by a 
superposition of two 
effects: the photoevaporation of the SE edge and the interaction between 
the two star-forming knots A and B.

\subsubsection{Zone 3}

\begin{figure*}
\centerline{\includegraphics*[width=\linewidth]{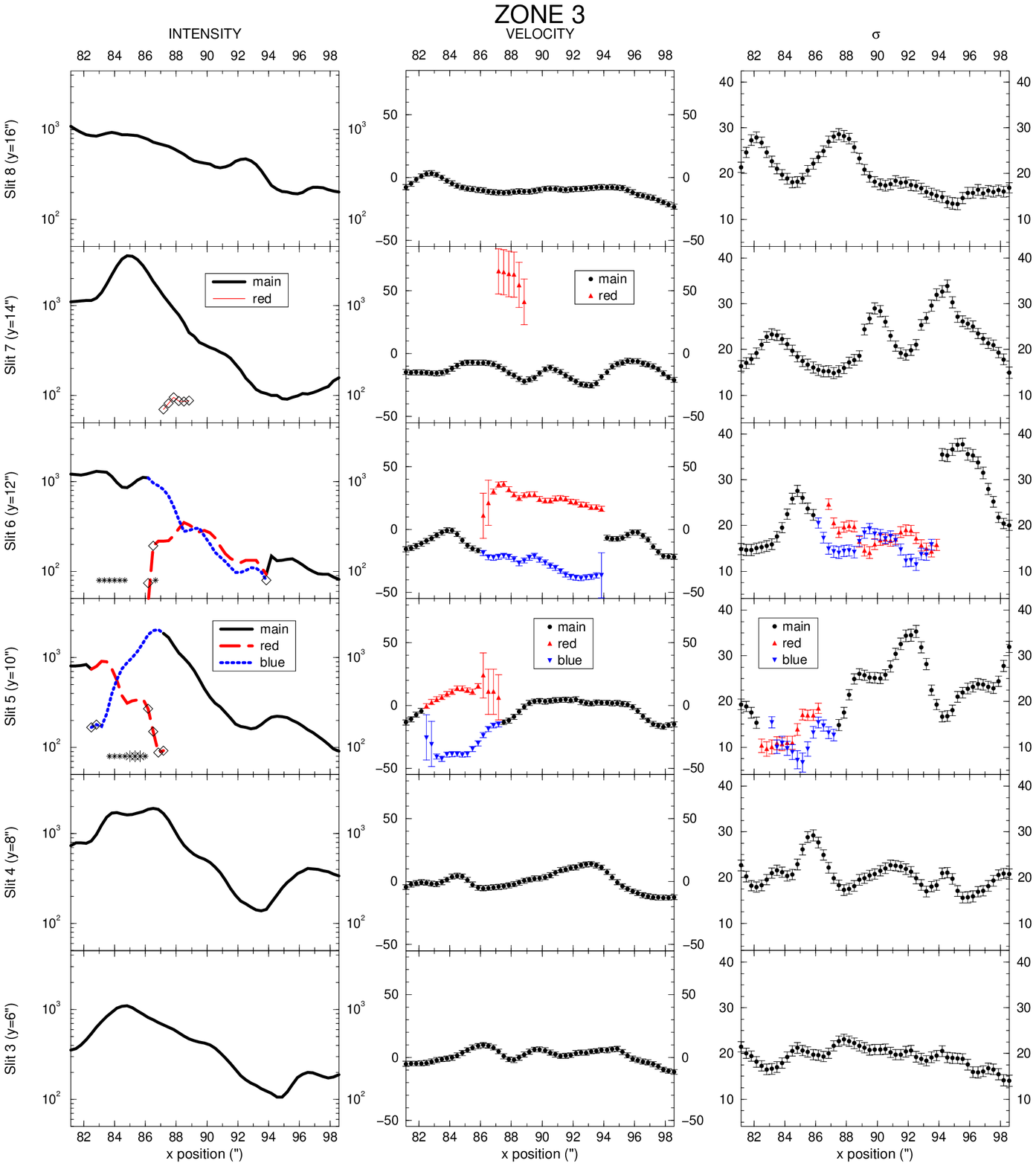}}
\caption{Intensity, velocity and \mbox{$\sigma$} values fitted to the H$\alpha$ 
emission 
corresponding to the spectra from zone 3. Symbols, lines, slit separation,
horizontal scale and units are as in Fig.~2. Asterisks correspond to continuum 
knot B. When two components are present and it is not possible to decide which
one is the main one, both are represented with a thick line. In these cases,
where possible, \mbox{$\sigma$} is shown for both components.}
\label{zone3}
\end{figure*}

\begin{figure}
\centerline{\includegraphics*[width=\linewidth]{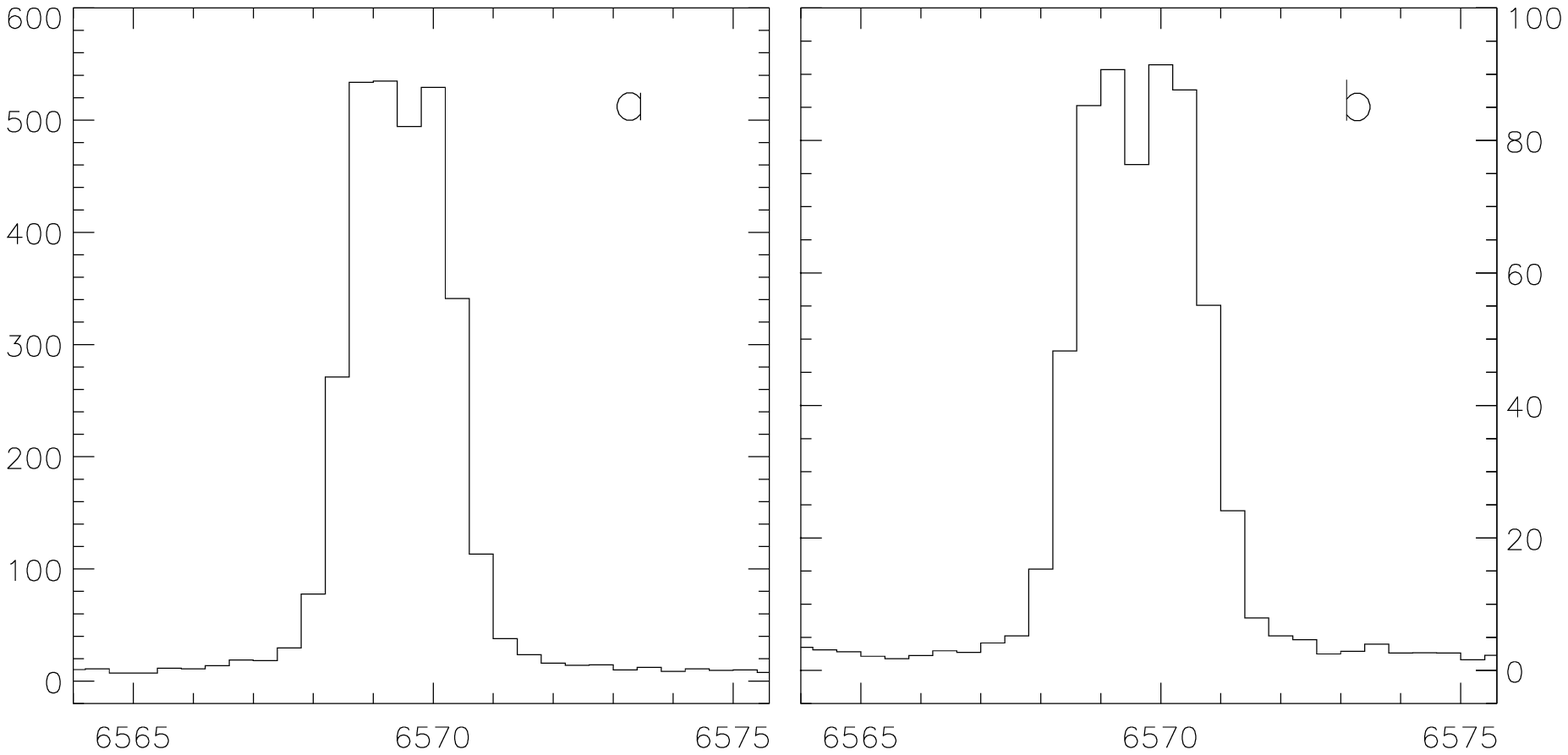}}
\caption{Spectra obtained at the two shells detected on zone 3. The spectrum on
the left (a) corresponds to the shell on slit 5 and the one on the right (b) to 
the one on slit 6. Units are the same as in Fig. 3.}

\label{spec_zone3}
\end{figure}

Zone 3 is traversed by 6 slits (numbers 3 to 8), and includes continuum
knot B (see Fig.~\ref{zone3}). The H$\alpha$ line parameters are shown in 
Fig.~\ref{zone3}. Coincident with the location of this continuum knot
(slit 5, $x$~=~85), an expanding shell with a velocity of about 25 km s$^{-1}$
can be seen.  The signature of this shell can be tracked both in the intensity
and velocity plots and is one of the few places where a main component cannot
be defined, i.e., two components of similar intensity are present at the same 
time. A spectrum is shown on Fig.~\ref{spec_zone3}a.  The
shell has a radius of 2.5\hbox{$^{\prime\prime}$} (50 pc) and is most probably 
broken as it
presents a red dominant component towards its northwestern end and a blue one 
in the southeastern direction.

As one moves to the right inside zone 3 the H$\alpha$ intensity drops by more 
than an order of magnitude, in a similar way to what happens in zone 2.  A
large-scale champagne flow has developed.  One can also see the signature of a
large expanding shell clearly reflected on the intensity and velocity plots 
(see Fig.~\ref{spec_zone3}b) of slit 6.  This shell is very elongated
and spans almost 10\hbox{$^{\prime\prime}$} (200 pc) along the density 
gradient, but its width
is smaller than 40 pc, as it only occurs in one slit.  Moving along the slit, 
the intensities of the red and the blue components add up to similar intensity
values displayed by the main component along either slits 5 or 7 at similar
locations along the density gradient.  The velocity separation implies an
expansion of some 25 km s\mbox{$^{-1}$} (see Fig.~\ref{zone3}). The $\sigma$ 
trend along 
the slit can also be understood as resulting from an expanding bubble. Each
component, when resolved, has the same line width as that of the gas nearby and
at the extreme ends of the slit.  The enhancements right at the boundaries of
the splitted line region can be understood as the integrated intensity along 
the line of sight of a thick shell which defines the boundary of the bubble.  
The width of the two bumps in the \mbox{$\sigma$} plot could be taken as a 
coarse measure 
of the shell thickness (around 5\hbox{$^{\prime\prime}$} or 100 pc).

The second bubble, extended along the density gradient, seems to be spatially 
connected to the smaller one (sitting on cluster B). Its most likely formation
mechanism is a blowout of the first superbubble, produced when evolving into 
the lower density medium (see Comer\'on 1997), as we will discuss later.

\subsubsection{Knot 11}

\begin{figure}
\centerline{\includegraphics*[width=\linewidth]{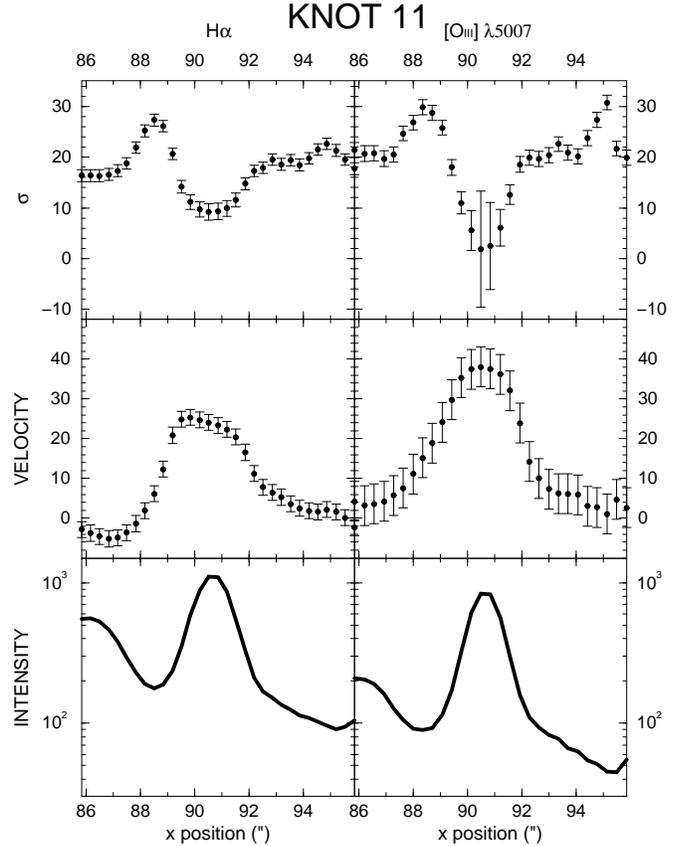}}
\caption{Intensity, velocity and \mbox{$\sigma$} values fitted to the H$\alpha$ 
emission 
and [O\,{\sc iii}]~$\lambda$5007 along the slit crossing knot 11 (slit number 
2). Symbols, lines, slit separation, horizontal scale and units are as in 
Fig.~2.}

\label{k11}
\end{figure}

To the NE border of zone 3, though clearly separated, there is a small (about
4\hbox{$^{\prime\prime}$} or 80 pc) emission knot (knot 11 on paper I and in 
Fig.~\ref{hacr0}).
This knot has been analyzed through a slit crossing the peak intensity maximum
(slit 2).  On each position of the slit, both H$\alpha$ and 
[O\,{\sc iii}]~$\lambda$5007 lines are well
fitted by a single Gaussian and the results are shown on Fig.~\ref{k11}.

Both emission lines show similar properties.  The physically small but intense
emitting knot displays a maximum and almost constant velocity of about 25
km s$^{-1}$ in the case of H$\alpha$ and 35 km s$^{-1}$ in the case of 
[O\,{\sc iii}]~$\lambda$5007 with 
respect to the systemic one. These velocities imply a redshift of some $25-30$ 
km s$^{-1}$ compared to its surroundings.  On both
sides of the intensity and velocity maxima the lines become broader by more
than 10 km s$^{-1}$ with respect to the nearby ISM.

\subsubsection{Zone 4}

Zone 4 represents the NE border (see Fig.~\ref{hacr0}) of NGC 4214-I and 
encloses a large area where the emission lines outside the important 
star-forming regions (clusters A and B) can be analyzed.  The zone contains the 
background ionized medium representative of the intercloud ISM of the galaxy.  

\begin{figure*}
\centerline{\includegraphics*[width=\linewidth]{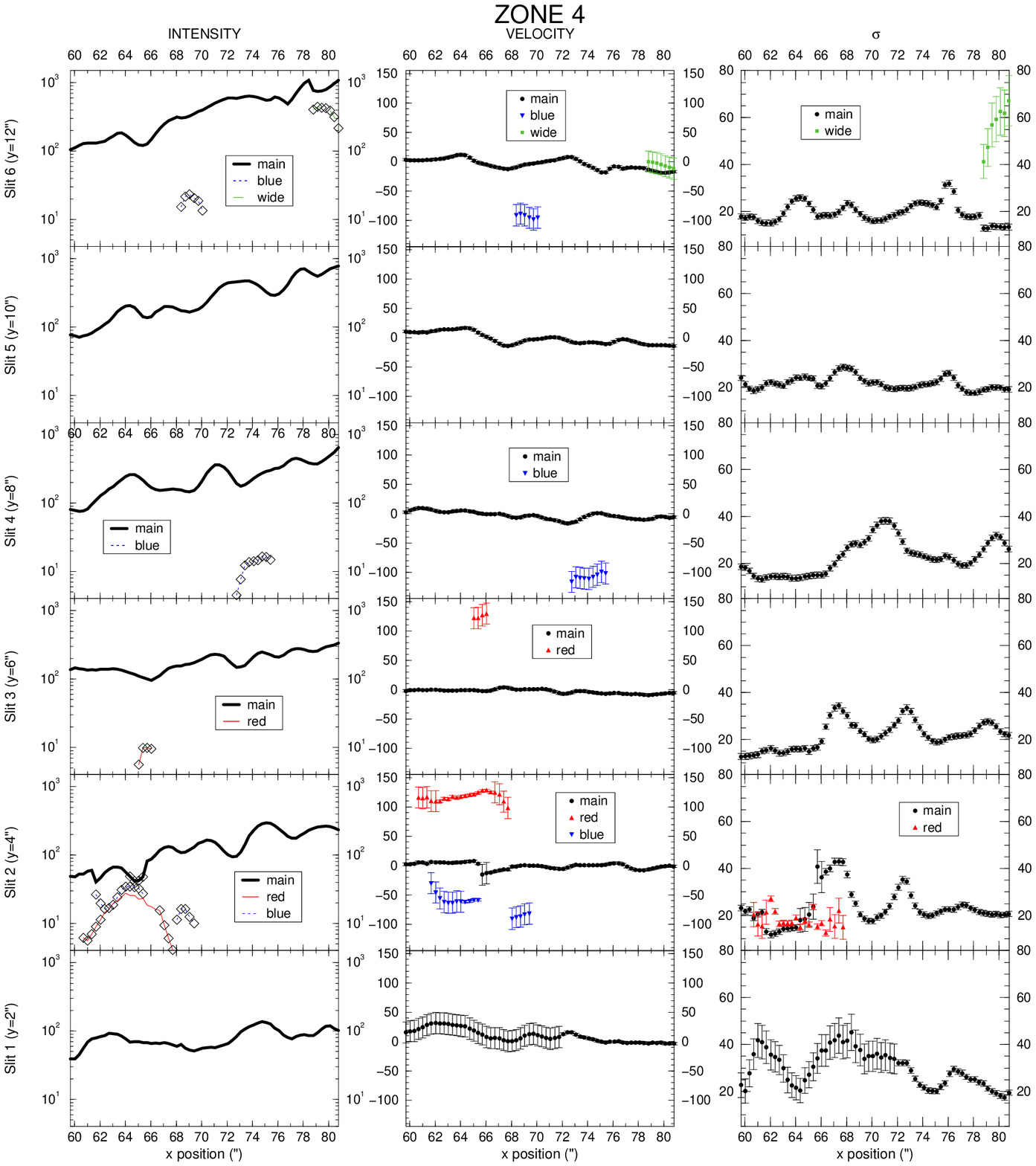}}
\caption{Intensity, velocity and \mbox{$\sigma$} values fitted to the H$\alpha$ 
emission corresponding to the spectra of zone 4. Symbols, lines, slit
separation, horizontal scale and units are as in Fig.~2.
Note that in this case we have been able to assign a 
\mbox{$\sigma$} value to a secondary component (compare with, e.g.,  
Fig.~\ref{zone1}), the wide component on slit 6.}

\label{zone4}
\end{figure*}

	As in the other three zones, a plot showing the intensity, 
velocity and $\sigma$ fits to 
the H$\alpha$ emission lines has been built and is presented in 
Fig.~\ref{zone4}.   The area is covered by 6 slits (slits 1 to 6) and spans 
from $x$~=~59\hbox{$^{\prime\prime}$} 
to $x$~=~81\hbox{$^{\prime\prime}$}.  Although zone 4 has been chosen to be far from the most
intense H$\alpha$ knots in zones 1 and 3, their presence can still be traced.  
As one moves to the southern corner of zone 4 the line intensity progresively
increases approaching that meassured on zones 1 and 3.  Therefore, there is not
a sharp boundary at NGC 4214-I as one moves towards the NE on the galaxy, but
instead a continuous extended medium all of it with the enhanced values of
$\sigma$ characteristic of the photoevaporation. The velocity of the main
component presents values close to the systemic velocity of NGC 4214, as 
expected,
considering the proximity of the area to the main star-forming knots of the
galaxy. A few secondary components are detected, including a wide one in slit 6
whose width (\mbox{$\sigma$} $\approx$ 60 km s\mbox{$^{-1}$}) could be 
measured. However, these 
components do not seem to possess an organized large scale structure, as
suggested by Martin~(1998). 

In zones 1, 2, 3 and 4, if one takes into account the differences in resolution
among the blue and red arms of the spectrograph, the 
[O\,{\sc iii}]~$\lambda$5007 emission behaves
in a very similar way as H$\alpha$.  However, the poorer spectral resolution 
of the spectrograph blue range has inhibited the measurement of the secondary
components.

\subsection{NGC 4214-II (the SE complex)}

\begin{figure*}
\centerline{\includegraphics*[width=\linewidth]{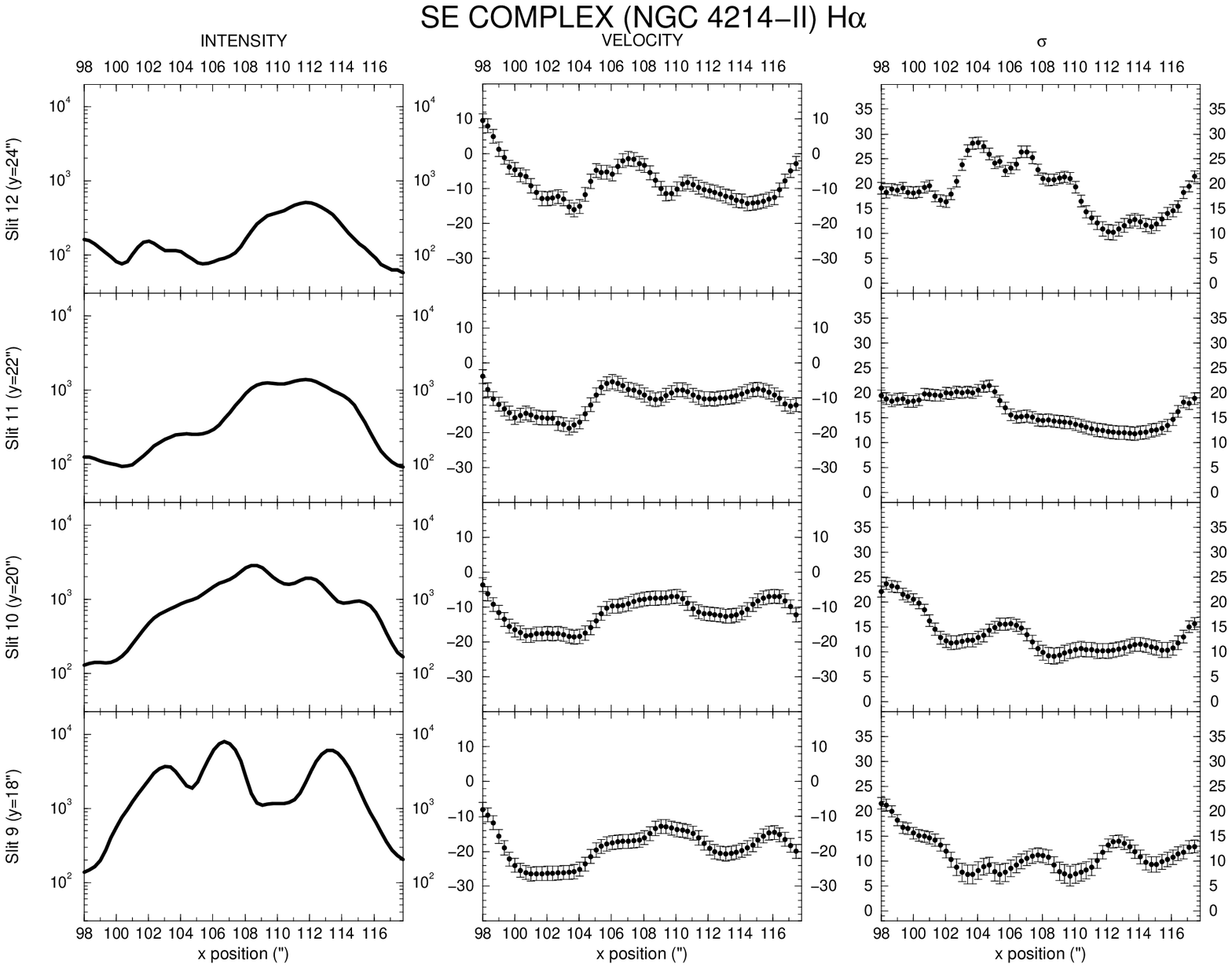}}
\caption{Intensity, velocity and \mbox{$\sigma$} values fit to the H$\alpha$ 
emission 
corresponding to spectra from zone SE. Symbols, lines, slit separation, 
horizontal scale and units are as in Fig.~2. The positions of continuum knots 
E, D and K are nearly coincident with the three intensity maxima on slit 9.}

\label{se}
\end{figure*}

\begin{figure*}
\centerline{\includegraphics*[width=\linewidth]{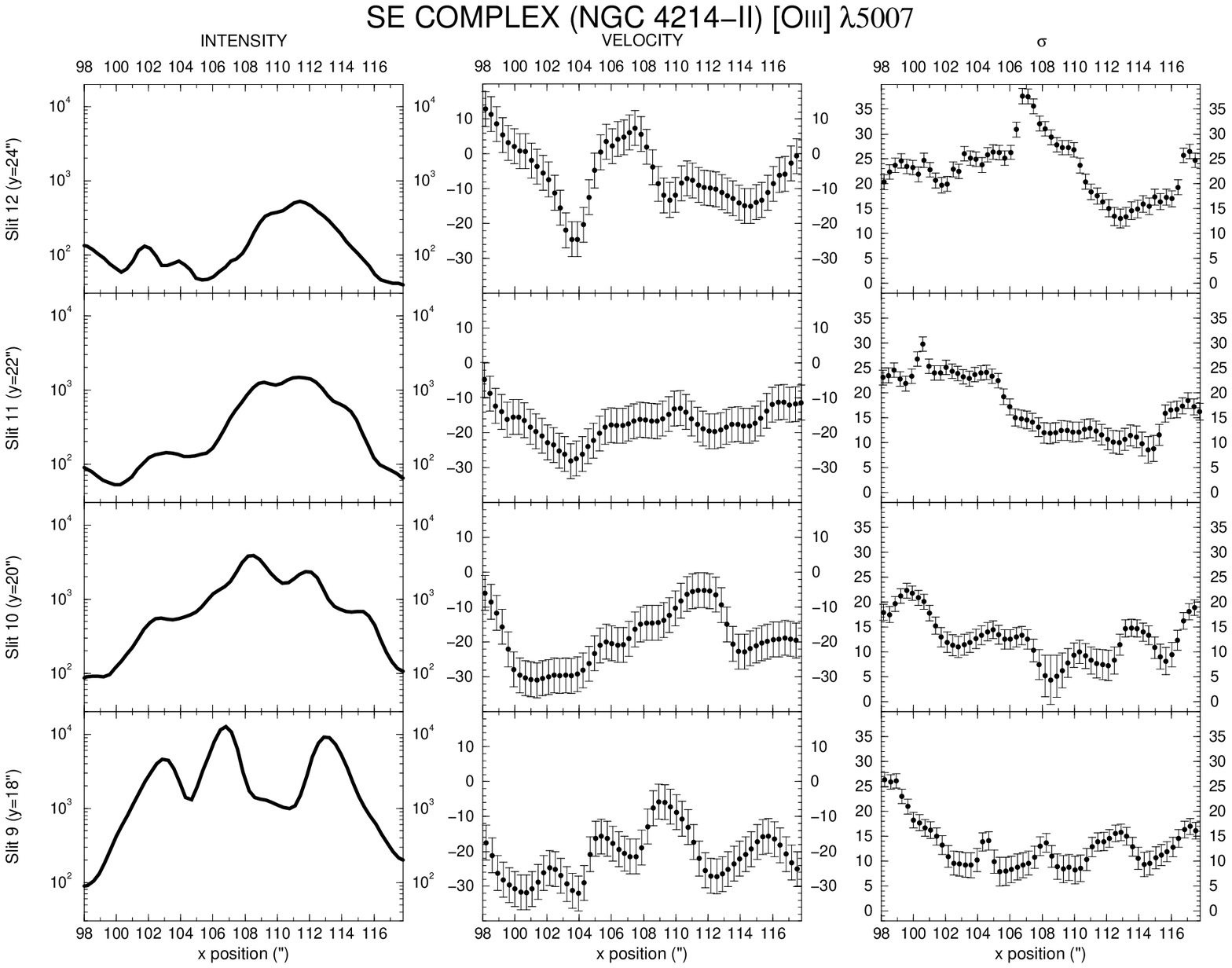}}
\caption{Intensity, velocity and \mbox{$\sigma$} values fit to the 
[O\,{\sc iii}]~$\lambda$5007 emission 
corresponding to spectra from zone SE. Symbols, lines, slit separation, 
horizontal scale and units are as in Fig.~2. The positions of continuum knots E,
D and K are nearly coincident with the three intensity maxima on slit 9.}

\label{seo}
\end{figure*}

The ionized gas emission of the star forming complex to the SE has been
analyzed through 4 slits starting with slit 9 which crosses the peak
emission maxima.  At each position on the slit, emission lines are
reasonably well described by one single Gaussian and results of the fit of
H$\alpha$ and [O\,{\sc iii}]~$\lambda$5007 are shown on 
Figs.~\ref{se}~and~\ref{seo}.  We start the
description with slit 9, which will be also taken as the reference for the
SE region.  The whole complex seems to be blueshifted about 10-20 km 
s\mbox{$^{-1}$}
with respect to NGC 4214-I.  Within the slit, the velocity spreads on a
range of about 15 km s\mbox{$^{-1}$}, implying small relative displacements of 
the ionized gas.  The velocity dispersion is around 10 km s\mbox{$^{-1}$} and, 
therefore,
sub-sonic almost all over the cloud, except for the left part of the region
(from $x$ = 98\hbox{$^{\prime\prime}$} to 101\hbox{$^{\prime\prime}$}). The 
\mbox{$\sigma$} value grows at both borders of
the complex, specially at the NW one.  On both extremes on the region, the
S/N is still good enough for a good fit to be done and artificial
broadening of the line due to this reason can be discarded.  This
possibility of photoevaporation taking place on the borders of the parent
cloud implies also a density bounded H\,{\sc ii} region and thus the three
intensity peaks could be local density enhancements close to the massive
star clusters and presently being disrupted by them.  As we move towards
the SW edge throughout slits 10, 11 and 12, the intensity decreases and the
identity of the individual knots is smeared out.  The velocity increases
reaching a maximum at position $x$ = 98\hbox{$^{\prime\prime}$} on slit 12. 
Line widths
also show a tendency to grow and some features showing local maxima both in
velocity and in $\sigma$ can be identified (see e.g., positions from
102\hbox{$^{\prime\prime}$} to 110\hbox{$^{\prime\prime}$} on slit 12). These 
again may be features showing
gas outflows from the density bounded regions. 

\subsection{The edge of the optical bar and the velocity curve}

On the right-hand side of Fig.~\ref{hacr0} we display other low intensity
features present along the optical bar of the galaxy.  Among these, the most
evident is the giant ring-like feature between $x$~=145\hbox{$^{\prime\prime}$} 
and 
$x$~=~165\hbox{$^{\prime\prime}$}, with a strong emitting knot (knot 16, in the 
nomenclature of 
Paper I).  Line widths of both red and blue components increase when 
reaching the intensity maximum at knot 16.
This giant ring could, at first glance, be interpreted as the result of a
large energy injection.  However, there is no trace of UV continuum associated
(Fanelli et al. 1997) and thus no trace of a stellar generation capable of 
blasting the ISM and causing the 400 pc diameter ring.

\begin{figure}
\centerline{\includegraphics*[width=\linewidth]{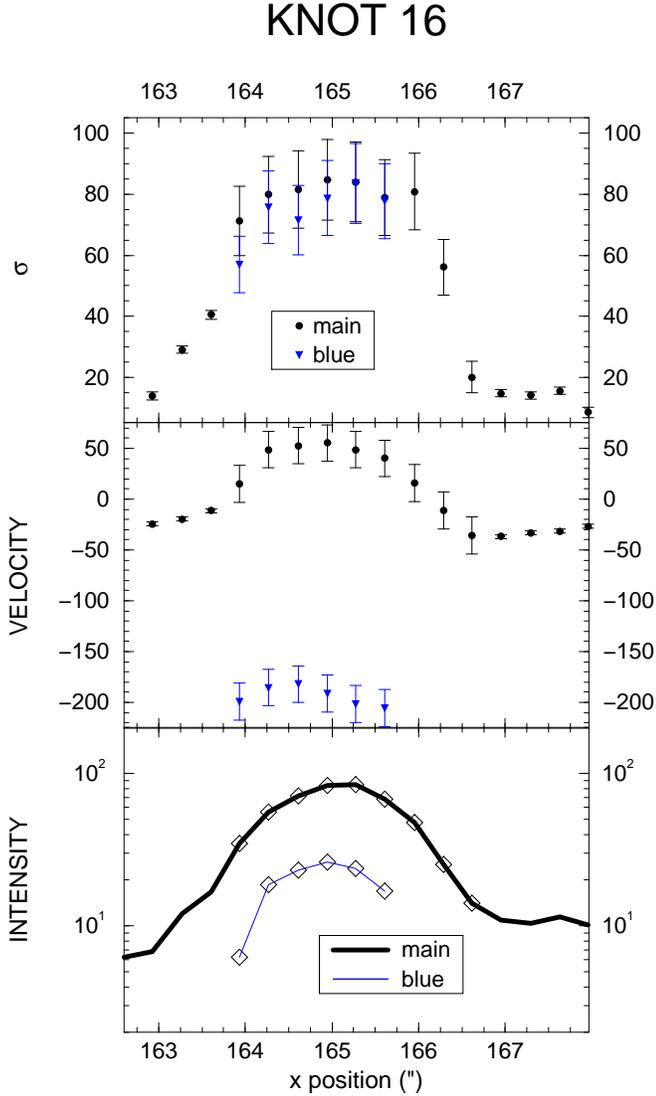}}
\caption{Intensity, velocity and \mbox{$\sigma$} values fitted to the H$\alpha$ 
emission along 
the slit crossing knot 16. Symbols, lines, slit separation, horizontal scale 
and units are as in Fig.~2.}

\label{k16}
\end{figure}

Knot 16 has been analyzed by fitting the emission lines along slit 8, which
traverses the peak intensity maximum.  Results of the H$\alpha$ fit are shown in
Fig.~\ref{k16}.  The emission line splits into two components across the knot,
and presents a main component an order of magnitude brighter than the
background.  The velocity of the most intense line reaches a maximum of 75 km
s\mbox{$^{-1}$}, compared with the systemic velocity measured in the nearby 
ISM.  The blue
line shows a higher relative velocity reaching up to $-$175 km s\mbox{$^{-1}$}.

\begin{figure*}
\centerline{\includegraphics*[width=\linewidth]{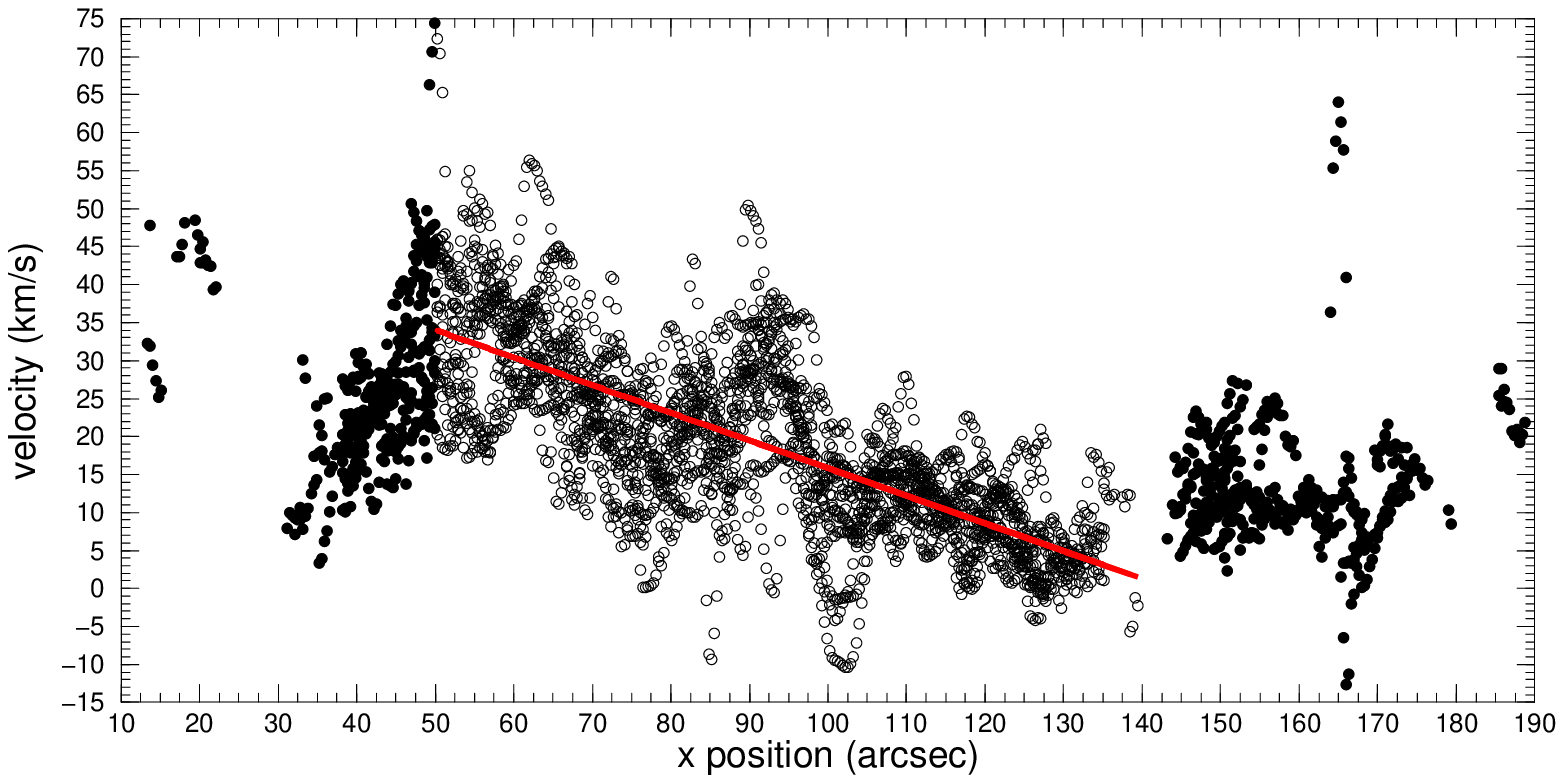}}
\caption{Rotation curve in H$\alpha$ along the bar of NGC~4214 compressed along 
the
direction perpendicular to the slit. The velocity corresponds only to the main 
(highest intensity) component in each pixel. Only those pixels with integrated 
intensity greater than 20$\cdot 10^{-17}$ erg s\mbox{$^{-1}$} cm\mbox{$^{-2}$} 
arcsec\mbox{$^{-2}$} are
included. The straight line is a linear fit for the points with 
50\hbox{$^{\prime\prime}$}$\le x \le 140$\hbox{$^{\prime\prime}$} (those 
represented as empty circles).}
\label{rotcurve}
\end{figure*}

Fig.~\ref{rotcurve} shows the rotation curve along the $x$ axis (i.e. parallel 
to the slits) derived from 
our data on the central regions of NGC~4214.  This was obtained from all the
pixels where the main H$\alpha$ component had a value above a threshold of 
20$\cdot 10^{-17}$ erg s\mbox{$^{-1}$} cm\mbox{$^{-2}$} arcsec\mbox{$^{-2}$}. 
The threshold was used in order 
to avoid values with low S/N ratio. Since twelve slits were used, that is the
maximum number of points that can appear for a given value of $x$.

The central 90\hbox{$^{\prime\prime}$} (1.8 kpc) can be reasonably well 
adjusted by a solid
body rotation with a projected velocity gradient of 18.2 km s\mbox{$^{-1}$} 
kpc\mbox{$^{-1}$} (assuming a
distance of 4.1 Mpc).  However, some pixels deviate from the mean rotation
curve by up to 30 km s$^{-1}$.  Three regions show strong deviations towards
the blue (around $x$~=~77\hbox{$^{\prime\prime}$}, 85\hbox{$^{\prime\prime}$},
and 101\hbox{$^{\prime\prime}$}) and another one
towards the red (around $x$~=~90\hbox{$^{\prime\prime}$}).  These regions 
correspond to
continuum knots A and B, and knot E (the NW part of the SE complex) for the
blueshifted cases and to knot 11 for the redshifted one.  In all cases, they
correspond to continuum maxima, indicating that the 
ionizing stellar clusters can cause peculiar motions in the ionized gas
that make them depart from the mean rotation curve determined across the bar.

The velocity curve along the $x$ direction coincides (within 5~km s$^{-1}$)
along the central 1.8 kpc with the H\,{\sc i} rotation curve of McIntyre (1998).
Although with much lower spatial resolution (30\hbox{$^{\prime\prime}$}), the 
H\,{\sc i} data extends
over a much larger area.  McIntyre also detected a much larger gradient along
the $y$ direction, which is expected since it corresponds approximately to the
minor axis of the galaxy.  That effect is not observed in our data.  We
attribute it to the short $y$ range in our data and to McIntyre lower spatial
resolution.

Outside the central 1.8 kpc, the rotation curve flattens to the NW and remains
approximately constant to the SE.  These results are also coherent with those
of McIntyre (1998), leading us to conclude that the rotation curves detected in
both hydrogen phases are quite similar.  The small detected departures can be
simply ascribed to random motions induced by the young stellar clusters.  A
similar conclusion was reached by Hartmann et al. (1986).

\section{Discussion}

The ISM around the nucleus of NGC 4214 is very complex and we do not
pretend to offer a complete explanation of its present morphology.  There
are, however, several features, in particular the largest structures and
detected flows, which offer the possibility of assessing the importance of
both photoionization and mechanical energy deposition by strong stellar
winds and SNe on structuring the ISM.

\subsection{The ionizing clusters of NGC 4214}

We list in Table 1 the observed and derived properties of the massive young 
ionizing clusters around the nucleus of NGC 4214 and their surrounding shells 
and bubbles.   In our sample we have included the two main ionizing clusters of 
the NW complex (A and B) and the corresponding three clusters of the SE complex
(E, D and K). As mentioned in Paper~I, these three clusters are studied
together, due to the impossibility of separating them with the available
data.  We have also included for comparison cluster C. 
The number of Lyman continuum photons ($N_{Lyc}$) has been derived from the
luminosity of the H$\beta$ line. The effective temperature ($T_{eff}$) of each 
cluster has been obtained from the line ratios given by Kobulnicky \& Skillman 
(1996) following the calibration of Cervi\~no \& Mas-Hesse (1994), as explained 
in Paper~I, except for knot C for which the prediction of the synthesis
models has been indicated. The observed radius ($R^o$) and expansion velocity
($v_{exp}^o$) for the superbubble(s) around each cluster are also shown in 
Table~1.  In the case of knot B, the first set
of observed values refers to the superbubble centered on the continuum knot
while the second set refers to the superbubble located towards the SE of
the continuum knot. We also indicate whether a blowout is observed and the
measured kinetic energy ($E_k^o$) contained in the observed
superbubble assuming that the emitting gas has a density of 100 
cm\mbox{$^{-3}$}.

The properties listed in Table 2 have been obtained by combining the 
observational measurements and derived parameters discussed above and in 
Paper~I with the predictions of evolutionary synthesis starburst models by 
Cervi\~no \& Mas-Hesse (1994) and Cervi\~no (1998), according to the following
procedure:

\begin{itemize}

\item Different parameters ($W($H$\beta)$, effective temperature, WR stars
presence,...) are used to constrain the age ($t_\star$) of the ionizing 
clusters. 

\item  Comparing the total continuum luminosity of each cluster at
H$\beta$ with the predictions of the models for the assumed age, we derive
the total mass of gas transformed into stars ($M_\star$) during the starburst 
episode. The model predictions are normalized to 1~$M_{\sun}$ of stars for an
instantaneous burst with a Salpeter Initial Mass Function between 2 and 
120~$M_{\sun}$. We have
extrapolated the derived mass of the clusters given in Table~2 to the range
0.1 to 120~$M_{\sun}$. 

\item The model predictions for the mechanical energy input rate (due to
stellar winds and supernovae) and the total release of mechanical energy
during the history of the cluster ($L_k$ and $E_k$ respectively) are
de-normalized by the total mass of the cluster at the given age.

\item Following Bodenheimer et al. (1979) we have finally computed the kinetic 
energy given to the gas upon photoionization during the champagne phase as
\begin{equation} E_{photo} = f k T_{eff} N_{Lyc} \end{equation}
where $f$ represents a correction factor that accounts for the fraction of
the photon energy that remains as heat and the energy radiated away and for
the fact that when the starburst is younger a higher number of ionizing
photons are emitted.  From the data given in Table~1 and assuming $f \sim$ 0.1 
(a somewhat arbitrary value), we derived the $E_{photo}$ values given in 
Table~2. 

\end{itemize}

\begin{table*} 

\caption[]{Observed and derived properties of the stellar clusters in NGC~4214  
and their surrounding bubbles.}

\begin{tabular}{@{\extracolsep{7mm}}lcccccc}
\hline 
Knot & $N_{Lyc}$      & $T_{eff}$   & $R^o$ & $v_{exp}^o$ & blowout & 
$E_k^o$       \\
     & $10^{51}$ s\mbox{$^{-1}$} & $10^3$ K & pc    & km s\mbox{$^{-1}$}     &         & 
$10^{50}$ erg \\
\hline 
A        &  4.5  & 37.0 & 100            & 70 & YES & 1.0 \\
B        &  1.8  & 37.0 & 50             & 25 & NO  & 0.9 \\
         &       &      & $200\times 40$ & 25 & YES & 0.6 \\
C        &  0.42 & 34.0 & 70?            &    & YES &     \\
E+D+K    &  4.4  & 38.5 & $<10$          &    & NO  &     \\
\hline
\end{tabular} 
%
%
\label{prop1}
\end{table*}

%

\subsection{The giant H II regions in NGC 4214}

The main characteristic of giant H\,{\sc ii} regions is their supersonic line
width (15 km s$^{-1}$ $\leq$ $\sigma$ $\leq$ 40 km s$^{-1}$), believed to
be related to the ionizing cluster formation stage (see Tenorio-Tagle
et al. 1993).  The action of stellar winds and supernova
explosions could also contribute to broaden the emission lines but, since
the $\sigma$ values correlate with the size of the emitting regions, as
shown by Terlevich \& Melnick (1981), the main driver seems to be related
to the original properties of the protocluster.  Perhaps the best criterium
to select the size of a giant H\,{\sc ii} region is through a determination of
the ``kinematic core'' (Mu\~noz-Tu\~n\'on et al., 1995).  This is the size
of the area of the ionized nebula across which a single Gaussian supersonic
line can be found.  Good examples of this are the giant H\,{\sc ii} regions of
M101 (Mu\~noz-Tu\~n\'on 1994) and those catalogued in NGC 4449
(Fuentes-Masip 1997).

NGC 4214-I presents supersonic emission line widths and has previously been
classified as a single giant H\,{\sc ii} region.  Arsenault \& Roy (1988), for
example, list it as a region 560 pc in diameter with an e-folding velocity
width of 21.1 km s$^{-1}$.  Our studies, however, (see Section 2 and Paper
I) have clearly shown that two well separated and massive young clusters
contribute to the ionization of NGC 4214-I (cluster A in zone 1 and B in
zone 3). The presence of two stellar clusters implies that NGC 4214-I may
not be a single giant H\,{\sc ii} region but rather the result of two
neighbouring ones.  Under such assumption one can infer dimensions and
velocity dispersion values for the two H\,{\sc ii} regions from our high
resolution data.  The H\,{\sc ii} region around cluster A has a radius 
comparable
to the distance between the cluster and the western edge of the dense
region, which presents the largest intensity and a supersonic $\sigma$
value ($\sigma$ = 17 km s$^{-1}$).  This radius is equal to 100 pc.  Note
that this also coincides with the cluster radius derived from HST data by
Leitherer et al. (1996). The region around cluster B seems to be more
distorted by the mechanical energy deposited by the massive stars.  This
H\,{\sc ii} region shows a central superbubble and a secondary larger bubble
extending along the edge of the cloud.  However, a radius and a $\sigma$
value can be obtained from the values of the still mechanically unperturbed
regions, or area exterior to the central shell.  When doing so we obtain
values of $R=100$ pc and $\sigma$ = 20 km s$^{-1}$. Although it is clear
that the uncertainties in the parameters associated to the two H\,{\sc ii}
regions are very large, the important result is to resolve the complex NGC
4214-I into two separate GHIIRs.  This new classification makes both of
them fall on the Terlevich \& Melnick (1981) $R$ vs. \mbox{$\sigma$} and 
L(H$\beta$)
vs. \mbox{$\sigma$} correlations. On the other hand, if we consider both as a 
single entity, the resulting region does not fall on the correlation.

The large H\,{\sc ii} region NGC 4214-II does not present a uniform supersonic 
line
width and, therefore, does not belong to the giant H\,{\sc ii} region class.

\subsection{Shells, bubbles, superbubbles}

We have presented kinematical evidence of the existence of three shells in
NGC~4214-I, one of them associated to the most massive starburst (cluster A
in zone 1) and two more some 400 pc to the south (associated to cluster B
in zone 3).  There is also a broken large-scale ring-like low intensity
feature (see Fig.~\ref{hacr0}, $x$ = 145\hbox{$^{\prime\prime}$} to
170\hbox{$^{\prime\prime}$}) located 1.8 kpc away from the nucleus.

\begin{figure}
\centerline{\includegraphics*[width=\linewidth]{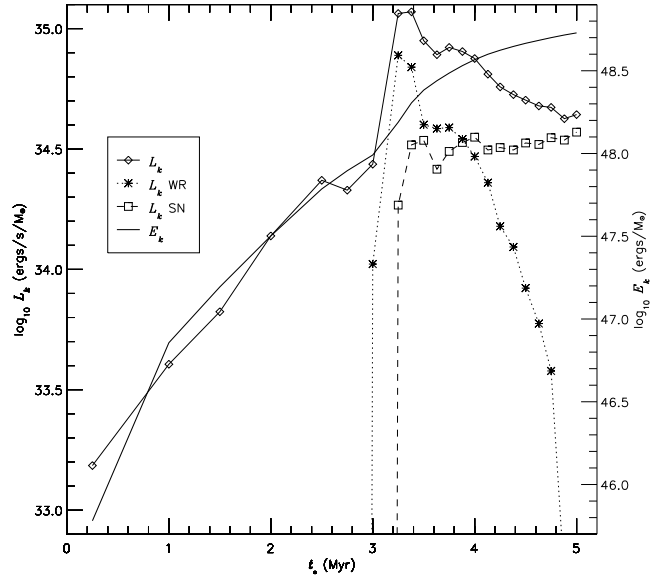}}
\caption{Mechanical energy input rate ($L_k$) and integrated value ($E_k$) per
unit mass as a function of time for an instantaneous burst with $Z=0.008$ 
(which is close to the metallicity of NGC 4214, see Kobulnicky \& Skillman
(1996)) and a Salpeter IMF between 2 and 120 M$_\odot$. $L_k$ is the sum of the 
contribution from winds from non-Wolf-Rayet stars, winds from Wolf-Rayet stars
and supernovae. The contribution to $L_k$ from Wolf-Rayet winds and supernovae 
has also been represented.}
\label{ek}
\end{figure}

Given the mechanical energy expected from the ionizing massive clusters and
their age, one can attempt to derive the properties of the superbubbles
(radius $R$, expansion speed $v_{exp}$, kinetic energy $E_k$) that they
should have produced, assuming a uniform mass density $\rho_0$ for the
background medium. This was done by Weaver et al. (1977) assuming a
constant mechanical energy input rate $L_k$.  However, that hypothesis is
not valid for a starburst. During the first 3 Myr after the formation of
the cluster, the mechanical energy input is dominated by winds from early
type O stars. These stars present an increasing wind luminosity with
time. This effect can be clearly seen in Fig.~\ref{ek}, where the
mechanical energy input rate and integrated value (obtained from the models
of Cervi\~no \& Mas-Hesse (1994) and Cervi\~no (1998)) as a function of
time have been plotted. Although the model of Weaver et al. (1977) is
therefore not directly applicable, an analytical solution can still be
found, as shown in the Appendix, yielding:

\begin{equation}
R(\mbox{kpc}) = 0.88 \left(\frac{L_{41}}{n_0}\right)^{\frac{1}{5}}
                     t_7^{\frac{3}{5}}
\label{Req4}
\end{equation}

\begin{equation}
v_{exp} (\mbox{km s\mbox{$^{-1}$}}) = 52 \left(\frac{L_{41}}{n_0}\right)^{\frac{1}{5}}
                              t_7^{-\frac{2}{5}},
\label{veq2}
\end{equation}

\noindent where $L_{41}$ is the present luminosity in units of $10^{41}$ erg
s\mbox{$^{-1}$} (not constant any more), $n_0$ is the background gas density 
in cm\mbox{$^{-3}$},
and $t_7$ is the evolution time in units of $10^7$ yr. This model has, of 
course, its limitations: As it can be seen in Fig.~\ref{ek}, a power law cannot
provide a good fit over periods of time longer than 3$-$4 Myr. However, it can
always serve as a comparison to the constant input rate model of Weaver et al. 
(1977) in order to provide an estimate of the dependence of $R$ and $v_{exp}$
on the variability of the mechanical energy input rate.

Applying Eqs.~\ref{Req4} and \ref{veq2} to the derived data of the different
regions we have obtained the size and expansion velocities of the expected
superbubbles. The results appear in Table~\ref{prop2}, where a background
density of 100 cm\mbox{$^{-3}$} has been assumed (see Paper I). These 
superbubble 
parameter predictions have to be taken with caution when compared with the
observational values. They would correspond to the properties that the
bubbles and shells should have assuming spherical symmetry and an
homogeneous interstellar medium.

\begin{table*} 

\caption[]{Some properties of the clusters and superbubbles in NGC~4214 obtained
from the synthesis models, as explained in the text. The assumed
background density is 100 cm\mbox{$^{-3}$}. $M_\star$ is normalized to the range 
$0.1-120$ M$_\odot$ assuming a Salpeter IMF (note that this normalization is
different from the one applied in Fig.~\ref{ek}). $R^t$ and $v_{exp}^t$ have
been calculated from Eqs.~\ref{Req4} and \ref{veq2}. However, since C is older 
than around 3.5 Myr, we have included not only the predictions of 
Eqs.~\ref{Req4} and \ref{veq2} but also those from Eqs.~\ref{Req5} and 
\ref{veq3}.}


\begin{tabular}{@{\extracolsep{7mm}}lccccccc}
\hline 
Knot     & $M_\star$ & $t_\star$ & $R^t$     & $v_{exp}^t$ & $L_k$      & $E_k$     & $E_{photo}$ \\ 
         & $10^5$ M$_\odot$ & Myr & pc      & km s\mbox{$^{-1}$}   & $10^{39}$ erg s\mbox{$^{-1}$}   
& $10^{51}$ erg       & $10^{51}$ erg         \\ 
\hline 
A        & 3.5 & 3.0-3.5 & 116     & 21        & 11.6       & 155       & 236         \\
B        & 1.0 & 3.0-3.5 & 91      & 17        & 3.5        & 46        & 94          \\
C        & 0.7 & 3.5-4.5 & 87-110  & 13-16     & 1.5        & 74        & 27          \\
E+D+K    & 1.6 & 2.8-3.3 & 70      & 14        & 1.2        & 44        & 213         \\
\hline
\end{tabular} 


\label{prop2}
\end{table*}

Around cluster A one can see the main emission from the galaxy across the
low intensity valley but not a complete shell of 116 pc across expanding
with a velocity of 21 km s$^{-1}$ as expected from the theory (see Table 2).  
The lack of a large kinematically detected shell around cluster A thus implies 
a limiting mechanism or a threshold value in the galaxy that inhibits the
build up of structures larger than about 100 pc in radius, implying the
blowout of the superbubble.  On the other hand, there is a ring-like
feature associated to the starburst located at the UV nucleus (see Section 
3.1.1), which presents
a red and a blue component in H$\alpha$ at various positions across it.  Both
components are shifted some 70 km s\mbox{$^{-1}$} from the main most intense 
line placed at a velocity $\sim v_{\rm galaxy}$.  The presence of this main
component implies that the stellar activity, even in the nucleus of the
galaxy, has not been efficient enough to sweep all the surrounding ISM into
a shell.  The shell-like structure is really a depression found in the
intensity of the various lines but it is not a cavity as that predicted by
theoretical models of stellar winds and SN remnants.  Furthermore, in a
velocity vs. position diagram the feature does not present the
characteristics typical of an expanding shell (such as the largest
splitting when observed towards the center and a single emission line
nearer the edge) and the observed $v_{exp}$ is much larger than the
expected value.  The feature, apart from the strongest emission centered at
the velocity of the galaxy, presents an extended red component spatially
shifted from the less extended blue counterpart.  The only geometry in
agreement with all these features is that of a young filled superbubble
bursting out on both sides of an almost face-on and thin galaxy disk and
undergoing a sort of jet-like process (Comer\'on 1997).  Lines of sight
across the bubble will display first the blue component (from the bursting
counterpart approaching us) and then a region of overlap between the blue
and the (now unobscured by the disk) red component.  The latter will extend
even further than the blue one when lines of sight begin to pick up the
part of the shell that is bursting out of the disk on our side but whose
furthest away side is expanding away from us. 

The superbubble expected around the star cluster B also seems to have burst
out of the parent cloud but this time into the intercloud medium causing an
elongated and larger secondary shell, both of them well detected in our
spectra (see Fig.~5).  The expanding shells are still detected; the first
one has experienced blowout whereas the second one is still complete and,
therefore, does not seem to have experienced blowout yet. It is interesting
to note that the expansion velocity expected is roughly consistent with the
observational value, despite the irregular geometry induced by the blowout of
the main shell. 

The above discussion is in agreement with recent observations reported by
Martin (1998) who finds evidences of shells and loops structures out of the
main optical body of NGC~4214.  She finds no peculiar kinematical patterns
associated to them.  This, together with the known fact that blowout has
already taken place on present day starbursts, points towards the presence
of secondary shells generated by sweeping an extended component of the 
H\,{\sc i} halo material.

For the three clusters in the SE complex (the youngest star-forming region
in NGC~4214), we find no peculiar kinematical features --except for the
champagne flows at the borders--, while from their evolutionary state and
associated energetics we would expect to find an $R \approx 70$ pc
superbubble at low expansion velocity ($\approx 14 $ km s\mbox{$^{-1}$}, see 
Table 2).  
This disagreement cannot be solved by assuming the existence of three smaller
unresolved bubbles around clusters E, D and K, since the expected radius in
that case would be around 60 pc for each one of them and should appear
resolved in our data.

One possible explanation to this apparent lack of a superbubble is the
existence of a strong density gradient. Such a gradient would have led to a 
blowout as described by Comer\'on (1997), similar to the one in the NW
complex. However, such a blowout should have produced effects which are not
observed in the SE complex: the maximum of the gas emission is not shifted
with respect to the continuum (as it is in the NW complex) and the H$\alpha$
profile does not deviate from a gaussian profile within the accuracy of our
data. Comer\'on (1997) predicts that asymmetries in the H$\alpha$ profile should
appear at a level of $\approx 0.03-0.05$ times the maximum intensity and we
do not find them even at a level 10 times lower. Therefore, the most likely
explanation for the non-detection of a superbubble in this region is its
real absence. A similar result is obtained for a small cluster in NGC~604
which is apparently younger than the rest of the recently formed stars
there (Ma\'{\i}z-Apell\'aniz et al. 1999).

What other possibilities can explain the non-existence of this superbubble in 
the SE complex?
We postulate that the origin is the existence of an inhibiting mechanism
for the creation of superbubbles by winds during the first 2-3 Myr after
the formation of the starburst. If we delay the onset of the formation of
the superbubble for such a period of time, both the non-detection of a
superbubble in the SE complex and the small size of the ones detected in
the NW complex could be explained. This delay with respect to the
theoretical predictions could be associated to the results of Bernasconi \&
Maeder (1996). Their calculations show that very massive ($> 60$
M$_{\sun}$) stars take 2 to 2.5 Myr to accrete their mass and that during
that period of time they are already burning hydrogen while hidden inside
their high density molecular gas cocoons. Therefore, the release of
mechanical energy by the most massive stars during the first 2~Myr after
the onset of a burst would be much smaller than expected, and so the
formation of bubbles and shells would not really start until the most
massive stars have completed their accretion processes, at a burst age of
around 2~Myr. We note here that due to this effect the $W($H$\beta)$ values
predicted by present evolutionary synthesis models, which assume all stars
at ZAMS at the same time, would be significantly overestimated. This could
explain the lack of detection of starbursts with $W($H$\beta)$ values above
300~\AA, while synthesis models predict values for young bursts well above
500~\AA.

Based on these results, we propose that NGC~4214 is a dwarf spiral galaxy,
with a disk much thinner than that of other irregulars with similar optical
appearance like, for example, NGC~4449 or NGC~2363. In this way all the
structures produced by the stellar mechanical energy input must become
disrupted after blowout, as soon as their remnants exceed the small
thickness of the galactic disk ($\sim$200 pc).  Once this happens and given
the orientation of the galaxy, there should remain no clear kinematical
information to be detected in the optical regime.  In such a case, it is
only the morphological detection of elongated rings, loops and holes that
can provide us with some idea about the power of the stellar energy input.
On the other hand, in irregular galaxies such as in NGC~4449, NGC~2366 or
Ho~II, with an ISM distributed into a much thicker disk, shell structures
can be larger, more massive, and longer lived.  In such a case remnants of
the stellar activity are more easily recognizable not only through their
morphology but also kinematically by means of well detected line splitting.
According to this scenario, the lack of well defined structures would not
be surprising when comparing NGC~4214 with, for example, NGC~4449
(Fuentes-Masip
et al. 1995) or with NGC~1569 (Tomita et al. 1994) for which similar
studies of the central 2 kpc with comparable, or even poorer, resolution
have yielded a large number of kinematical features and thus to what at
first glance may seem a richer structure of the ISM.

This scenario would also be consistent with the overestimation by
evolutionary synthesis models of the total mechanical energy released, as
it is clear from Tables 1 and 2 when comparing the predicted and the
measured values. First of all, the delay in the apparition of the most
massive stars would significantly reduce the total release of mechanical
energy compared with the prediction of synthesis models. Furthermore, the
disruption of the structures discussed above inhibits the measurement of the
kinetic energy of the gas in the optical regime.  And, finally, an important
fraction of the mechanical energy should have contributed to heat the
interstellar gas, which would cool by the diffuse X-ray emission
detected in these galaxies, as discussed in Cervi\~no (1998).

The only large-scale ring structure detected is the one at 1.8~kpc from the
nucleus.  As already discussed on Sect.~3 it is unlikely that the structure
had been generated by present day star formation. It is most surely
produced by the general UV background of the galaxy.  The feature is also
identified in the H\,{\sc i} maps of McIntyre (1998), where it presents also a
ring-like shape around a central depression.  It therefore seems very likely
that the feature results from the partial ionization of the H\,{\sc i} ring
bordering the disrupted and yet undetected giant molecular cloud being
eroded by the background UV flux.  The giant cloud would sit at the end of
the bar where matter accumulation is probably leading to the build up of
clouds.  In such a scenario, the giant cloud is thus predicted to fill the
central depression seen in the H\,{\sc i} distribution.

\subsection{Photoionization champagne flows and froth features} 
   
NGC 4214-I might be density bounded, according to Leitherer et al. (1996)
and Sargent and Fillipenko (1991). We found in Paper~I that most of the
Lyman continuum photons produced by the young stellar clusters were
ionizing the dense gas clouds located to the south of them. The
inhomogeneous distribution of gas suggests that some fraction of these
photons might be leaking from the star-forming regions.  This might also be the
case for all the H\,{\sc ii} regions in the SE complex.  Along the borders of 
all of them, sharp line intensity drops of more than an order of magnitude are
accompanied by a sudden increase in the gas velocity dispersion up to
supersonic speeds.

Both of these flow features agree with the predictions from the champagne
phase (Tenorio-Tagle 1979).  UV photons leaking out of a dense cloud into
the surrounding medium generate upon photoionization a large pressure
difference causing the champagne flow and with it the dispersal of the
dense cloud.  Therefore, density bounded H\,{\sc ii} regions here and in any
other galaxy must display larger values of \mbox{$\sigma$} across their borders,
being this \mbox{$\sigma$} value provided by the matter expelled at supersonic 
speeds from the ionized cloud into the background during this champagne phase.
Following Bodenheimer et al. (1979) one can calculate the kinetic energy
given to the gas upon photoionization during the champagne phase by
Eq.~2.  As explained above, we have assumed $f \approx 0.1$, but this
value could be smaller if some of the UV photons are not absorbed across
the champagne flows and escape the main disk of the galaxy.  In any case,
the values of $E_{photo}$ listed in Table~2 would be sufficient to justify
the large $\sigma$ values observed at the edges of all ionized clouds.

The implication of such energetics has a great impact on the parent
clouds which might become disrupted in a time scale of the order of a few
times the crossing time ($t_{ch}$ $\sim$ cloud size/$a c_{\rm H\, II}$,
where $c_{\rm H\, II}$ is the speed of sound for an H\,{\sc ii} region and $a$ 
is a small number in the range 3 to 5).

In NGC 4214-II the outflow can be traced some 200 pc away from the sharp
intensity (density) discontinuity and covers the region where Hunter \&
Gallagher (1990) detected some ``froth'' features.  The outflow away from
NGC 4214-I presents much larger velocities and thus it is very likely that a
large fraction of the energy of the nuclear starburst is presently being
vented into the halo of the galaxy.  All of these features in the SE and NW
complexes, together with the size of the main superbubble point at a very
narrow galaxy disk of at most 200 pc across.  The disk is
presently being extended or lifted through champagne flows causing a
diffuse broad (200 pc) layer on top of the main galaxy disk.  The
interaction of neighbouring champagne flows would likely lead to sharp, and
more intense regions at the planes of interaction that could more easily be
detected and we believe could then be recognized as froth.

We thus conclude that the mechanical input deposited by the star formation
complexes, in a variety of physical processes that include the free
expanding bubbles liberated after blowout and photoevaporation of the
parent clouds, have succeeded in generating structures now detected far from
the disk.  The latter also supports the presence of and extended H\,{\sc i} 
halo,
sometimes partly ionized and seen as the DIG of the galaxy.  The 
impact of starbursts gives place to the large-scale structure which now
enriches the optical appearance of the galaxy.

\section{Conclusions}

We have analyzed the kinematical properties of the ionized gas around the
massive star clusters in the nucleus of NGC~4214 which were studied in
Paper~I. The main results we have obtained can be summarized as
follows: 

\begin{itemize}

\item Emission lines splitted into several components, as well as
significant variations of the line widths, have been detected in several
places. 

\item The Giant H\,{\sc ii} region around the two most massive clusters in
NGC~4214 (A and B) is resolved into two clearly separated regions, each one of
them associated to one of the clusters. Both H\,{\sc ii} regions taken 
separately fall
on the Terlevich \& Melnick (1981) $R$ vs. \mbox{$\sigma$} and L(H$\beta$) vs. 
\mbox{$\sigma$}
correlations. On the other hand, if they are taken as a single region they 
do not fall on the correlations.

\item There are several other young star clusters (C, D, E and K among them)
which are not massive enough to produce a Giant H\,{\sc ii} region around
them. Instead, the result is a regular H\,{\sc ii} region.

\item We have not detected superbubbles with the properties that would be 
expected
according to the evolutionary state of the stellar clusters. Around cluster
A, only a ring-like feature with three kinematical components at various 
positions has been
identified. Around cluster B, two expanding shells have been identified;
the first one seems to have experienced blowout, whereas the second one is
still complete. No kinematical features indicating the presence of shells
and bubbles have been found around the clusters of the SE complex. These 
results are in agreement with the Bernasconi \& Maeder (1996) model, in which
massive stars spend their first 2-3 Myr inside their cocoons.

\item We postulate that NGC~4214 is a dwarf spiral galaxy, with a
thin ($\sim 200$ pc) disk that inhibits the formation of large scale
structures in the ISM, which should have been disrupted after blowout, as
soon as their remmants exceed the galactic disk.

\item Champagne flows might have formed at the borders of the regions,
especially on the SE complex, explaining the existence of the diffuse
ionized gas around the galaxy.

\end{itemize}

\begin{acknowledgements}

The authors would like to thank the referee, C. Robert, for her useful advice
on how to improve this paper.
JMA would like to acknowledge the hospitality of the Instituto de
Astrof\'{\i}sica de Canarias and the Royal Greenwich Observatory, where
part of this work has been carried out. This work is based on observations
taken with the William Herschel Telescope at El Roque de los Muchachos
Observatory located on the island of La Palma. The observations were obtained
within the GEFE ({\sl Grupo de Estudios de Formaci\'on Estelar}) collaboration,
an international group whose main objective is the understanding of the 
parameters that control massive star formation in starbursts. This work has 
been partially supported by DGICYT grant PB94-1106, by CICYT grant 
ESP95-0389-C02-02, and by the IAC.

\end{acknowledgements}

\appendix
\section*{Appendix}

	In this appendix we discuss an analytical solution for the expansion of
an adiabatic wind-driven bubble in a uniform background medium with mass 
density $\rho_0$. Weaver et al. (1977) developed a model for the case where the 
bubble was powered by a single star with constant mechanical energy input rate 
$L_k$. In that case, if $E$ is the internal energy, $R$ is the radius of the 
bubble and $p$ is the pressure, the equations to be applied (the ideal gas
equation of state and the momentum and energy balance equations) can be
expressed as:

\begin{equation}
E=2\pi R^3 p
\label{eqstate}
\end{equation}

\begin{equation}
\frac{d}{dt}\left(R^3\rho_0\frac{dR}{dt}\right)=3R^2p
\label{mombal}
\end{equation}

\begin{equation}
\frac{dE}{dt}=L_k-4\pi R^2p\frac{dR}{dt}.
\label{enbal}
\end{equation}

        The above equations have an analytical solution which can be expressed
as:

\begin{equation}
E=\frac{5}{11}L_kt
\end{equation}

\begin{equation}
R=\left(\frac{250}{308\pi}\right)^{\frac{1}{5}}L_k^{\frac{1}{5}}
  \rho_0^{-\frac{1}{5}}t^{\frac{3}{5}}
\label{Req1}
\end{equation}

\begin{equation}
p=\frac{7}{\left(3850\pi\right)^{\frac{2}{5}}}L_k^{\frac{2}{5}}
  \rho_0^{\frac{3}{5}}t^{-\frac{4}{5}}.
\end{equation}

        Here we are interested in the evolution of two kinematical quantities:
$R$ and the expansion velocity $v_{exp}=\frac{dR}{dt}$. In more convenient
units, those are:

\begin{equation}
R(\mbox{kpc}) = 1.1 \left(\frac{L_{41}}{n_0}\right)^{\frac{1}{5}}
                    t_7^{\frac{3}{5}}
\label{Req5}
\end{equation}

\begin{equation}
v_{exp} (\mbox{km s\mbox{$^{-1}$}}) = 66 \left(\frac{L_{41}}{n_0}\right)^{\frac{1}{5}}
                              t_7^{-\frac{2}{5}}
\label{veq3}
\end{equation}

\noindent where $L_{41}$ = $L_k$/10$^{41}$erg s$^{-1}$, $t_7$ is the
evolution time in units of 10$^7$yr and $n_0$ is the density of the
background gas in cm$^{-3}$.

	The solution to be developed here has a power-law dependence for the 
mechanical energy input rate $L_k=At^\alpha$, where $A$ and $\alpha$ are 
constants. In that case, Eqs. \ref{eqstate}, \ref{mombal} and \ref{enbal} can 
still be solved analytically, the result being:

\begin{equation}
\begin{array}{lllll}
E & = & \frac{5}{11+3\alpha}At^{1+\alpha}
  & = & \frac{5}{11+3\alpha}L_kt
\end{array}
\end{equation}

\begin{equation}
\begin{array}{lll}
R & = &
  \left(\frac{375}{2\pi(3+\alpha)(7+4\alpha)(11+3\alpha)}\right)^{\frac{1}{5}}
  A^{\frac{1}{5}}\rho_0^{-\frac{1}{5}}t^{\frac{3+\alpha}{5}} \\
  & = &
  \left(\frac{375}{2\pi(3+\alpha)(7+4\alpha)(11+3\alpha)}\right)^{\frac{1}{5}}
  L_k^{\frac{1}{5}}\rho_0^{-\frac{1}{5}}t^{\frac{3}{5}}
\end{array}
\label{Req2}
\end{equation}

\begin{equation}
\begin{array}{lll}
p & = & \left(\frac{(3+\alpha)^3(7+4\alpha)^3}
                   {67500\pi^2(11+3\alpha)^2}\right)^{\frac{1}{5}}
        A^{\frac{2}{5}}\rho_0^{\frac{3}{5}}t^{\frac{2\alpha -4}{5}} \\
  & = & \left(\frac{(3+\alpha)^3(7+4\alpha)^3}
                   {67500\pi^2(11+3\alpha)^2}\right)^{\frac{1}{5}}
        L_k^{\frac{2}{5}}\rho_0^{\frac{3}{5}}t^{-\frac{4}{5}}.
\end{array}
\end{equation}

        As it can be seen, when the radius is expressed as a function of the
present luminosity $L_k$, the only difference between
Eqs.~\ref{Req1}~and~\ref{Req2} is the constant which multiplies
$L_k^{\frac{1}{5}}\rho_0^{-\frac{1}{5}}t^{\frac{3}{5}}$.

        Applying a least-squares fit to the evolution of the mechanical energy
input  with time (see Fig.~\ref{ek}), it can be seen that $L_k$  can be 
reasonably well adjusted by a power law during the first three million years
after the formation of the starburst. The best fit yields
$\alpha = 1.25$. In that case, the constant in Eq.~\ref{Req2} is lower by
a factor of 0.79 than the one in Eq.~\ref{Req1}. Therefore, for the case
$\alpha = 1.25$ we have:

\begin{equation}
R(\mbox{kpc}) = 0.88 \left(\frac{L_{41}}{n_0}\right)^{\frac{1}{5}}
                     t_7^{\frac{3}{5}}
\label{Req3}
\end{equation}

\begin{equation}
v_{exp} (\mbox{km s\mbox{$^{-1}$}}) = 52 \left(\frac{L_{41}}{n_0}\right)^{\frac{1}{5}}
                              t_7^{-\frac{2}{5}} .
\label{veq1}
\end{equation}


\begin{thebibliography}{}

\bibitem{ArseRoy86}
 Arsenault, R., Roy, J.-R., 1988, A\&A, 201, 199 

\bibitem{B} 
 Bodenheimer, P., Tenorio-Tagle, G., Yorke, H. W., 1979, ApJ, 233, 85

\bibitem{BM} 
 Bernasconi, P. A., Maeder, A., 1996, A\&A, 307, 829

\bibitem{CMH} 
 Cervi\~no M., Mas-Hesse J. M., 1994,   A\&A,   284, 789

\bibitem{Cer} 
 Cervi\~no M., 1998, PhD thesis, Universidad Complutense, Madrid

\bibitem{Com}
Comer\'on, F., 1997, A\&A, 326, 1195


\bibitem{Fan} 
 Fanelli, M. N. et al. 1997, ApJ, 481, 735

\bibitem{FMetal1}
 Fuentes-Masip, O., Casta\~neda, H. O., Mu\~noz-Tu\~n\'on, C., 1995, A.S. P. 
 Conference Series, 71, 143.

\bibitem{FM}
 Fuentes-Masip, O., 1997, PhD., Universidad de La Laguna

 \bibitem{HGH} 
 Hartmann L.W., Geller M.J., Huchra J. P., 1986,   AJ, 92, 1278

\bibitem{HG}
 Hunter, D. A., Gallagher, J. S. III, 1990,   ApJ,   362, 480

\bibitem{ks96} 
 Kobulnicky, H. A., Skillman, E. D., 1996, ApJ, 471, 211 

\bibitem{Leitetal} 
 Leitherer C. Vacca, W. Conti, P. S., Filippenko A. V.,
 Robert C., Sargent W. L. W., 1996,   ApJ, 465, 717 

\bibitem{Maiz1}
 Ma\'{\i}z-Apell\'aniz, J., Mas-Hesse, J. M., Mu\~noz-Tu\~n\'on, C., 
 V\'{\i}lchez, J. M., Casta\~neda, H. O., 1998, A\&A, 329, 409 (Paper I)

\bibitem{Maiz2}
 Ma\'{\i}z-Apell\'aniz, J. et al. , 1999, in preparation 

\bibitem{Martin}
 Martin, C.L., 1998, ApJ, in press.



\bibitem{Mc}
 McIntyre, V. J., 1998, PASA, 15, 157.

\bibitem{cmt94}
 Mu\~noz-Tu\~n\'on, C., 1994, in {\it Violent Star Formation:
 From 30 Doradus to QSOs}, ed. G. Tenorio-Tagle,
 Cambridge University Press, 25

\bibitem{cmt95}
 Mu\~noz-Tu\~n\'on, C., Gavryusev, V., Casta\~neda, H. O., 1995, 
 AJ, 110, 1630

\bibitem{cmt97}
 Mu\~noz-Tu\~n\'on, C., Fuentes-Masip, O., Casta\~neda, H. O., 1998, PASA, 15, 
 103.

\bibitem{roy86}
 Roy, J.-R., Arsenault, R., Joncas, G., 1986, ApJ, 300, 624 

\bibitem{} Sandage A., Bedke J. 1985,   AJ,   90, 1992

\bibitem{sf91} 
 Sargent, W. L. W., Filippenko, A. V., 1991,   AJ,   102, 107 

\bibitem{Ten}
 Tenorio-Tagle, G., 1979, A\&A, 71, 59

\bibitem{TMTC}
 Tenorio-Tagle, G., Mu\~noz-Tu\~n\'on, C., Cox, D. P., 1993, ApJ, 418, 767 


\bibitem{TMTCF2}
 Tenorio-Tagle, G., Mu\~noz-Tu\~n\'on, C., P\'erez, E., Melnick, J., 1997,
  ApJ, 490, L179.
 
\bibitem{TM}
 Terlevich, R., Melnick, J., 1981, MNRAS, 195, 839

\bibitem{Tom}
 Tomita, A., Ohta, K., Saito, M., 1994, PASJ, 46, 335

\bibitem{Weaver}
Weaver, R. et al. , 1977, ApJ, 218, 377.
 
\end{thebibliography}
\end{document}